\documentclass[longauth]{aa}

\usepackage{graphicx}	% Including figure files
\usepackage{amsmath}	% Advanced maths commands
\usepackage{amssymb}	% Extra maths symbols
\usepackage{multicol}	% Multi-column entries in tables
\usepackage{bm}		% Bold maths symbols, including upright Greek
\usepackage{pdflscape}	% Landscape pages
\usepackage{ulem} %strike out text
\usepackage{multirow}
\usepackage{caption}

\usepackage[usenames,dvipsnames]{xcolor}

\usepackage[draft,hyperfootnotes=false]{hyperref}
\hypersetup{
  colorlinks,
  citecolor=Blue,
  linkcolor=Blue,
  urlcolor=Blue}  

\usepackage{xspace}

\usepackage{verbatim}
\usepackage{tabulary}
\usepackage{tabularx}
\usepackage{todonotes}
\usepackage{cprotect}
\usepackage{newtxtext}

\let\oldpageref\pageref
\renewcommand{\pageref}{\oldpageref*}

\newcommand{\Omegam}{\Omega_{{\rm m}}}
\newcommand{\Omegab}{\Omega_{{\rm b}}}
\newcommand{\OmegaL}{\Omega_{{\Lambda}}}

\newcommand{\nv}{\hat{\bf n}}
\newcommand{\var}[1]{\ensuremath{\texttt{\MakeUppercase{#1}}}\xspace}
\newcommand\code[1]{\texttt{\small{#1}}}

\newcommand{\gold}{\code{Y3\,GOLD}\xspace}
\newcommand{\sample}{{\rm {BAO\,sample}}\xspace}

\newcommand{\photoz}{photo-$z$\xspace}

\newcommand{\DNF}{\code{DNF}\xspace}

\newcommand{\ZMEAN}{\code{Z\_MEAN}\xspace}
\newcommand{\ZMC}{\code{Z\_MC}\xspace}

\newcommand\cosmolike{{\textsc{CosmoLike} }}
\newcommand{\LCDM}{$\Lambda$CDM }

\usepackage{eso-pic}% http://ctan.org/pkg/eso-pic

\AddToShipoutPictureBG*{%
  \AtPageUpperLeft{%
    \hspace{0.75\paperwidth}%
    \raisebox{-1.5\baselineskip}{%
      \makebox[0pt][l]{\textnormal{DES-2019-0518}}
}}}%

\AddToShipoutPictureBG*{%
  \AtPageUpperLeft{%
    \hspace{0.75\paperwidth}%
    \raisebox{-2.5\baselineskip}{%
      \makebox[0pt][l]{\textnormal{FERMILAB-PUB-21-311-AE}}
}}}%

\usepackage{txfonts}

\begin{document}
% Don't change these lines

    \titlerunning{DES Y3 BAO mock catalogs}
    \title{Dark Energy Survey Year 3 Results: Galaxy mock catalogs for BAO analysis}
    
\author{
\href{https://orcid.org/0000-0002-1295-1132}{I.~Ferrero}\inst{1,2,3,}\thanks{s.i.ferrero@astro.uio.no} \and
{M.~Crocce}\inst{2,3} \and
{I.~Tutusaus}\inst{2,3} \and
{A.~Porredon}\inst{4,5} \and
{L.~Blot}\inst{6} \and
{P.~Fosalba}\inst{2,3} \and
{A.~Carnero~Rosell}\inst{7,8,9} \and
{S.~Avila}\inst{10} \and
{A.~Izard}\inst{11} \and
{J.~Elvin-Poole}\inst{4,5} \and
{K. ~C.~Chan}\inst{12} \and
{H.~Camacho}\inst{13,8} \and
{R.~Rosenfeld}\inst{14,8} \and
{E.~Sanchez}\inst{15} \and
{P.~Tallada-Cresp\'i}\inst{15,16} \and
{J.~Carretero}\inst{17,16} \and
{I.~Sevilla-Noarbe}\inst{15} \and
{E.~Gaztanaga}\inst{2,3} \and
{F.~Andrade-Oliveira}\inst{13,8} \and
{J.~De~Vicente}\inst{15} \and
{J.~Mena-Fern{\'a}ndez}\inst{15} \and
{A.~J.~Ross}\inst{4} \and
{D.~Sanchez Cid}\inst{15} \and
{A.~Fert\'e}\inst{18} \and
{A.~Brandao-Souza}\inst{19,8} \and
{X.~Fang}\inst{20} \and
{E.~Krause}\inst{20} \and
{D.~Gomes}\inst{21,22} \and
{M.~Aguena}\inst{8} \and
{S.~Allam}\inst{23} \and
{J.~Annis}\inst{23} \and
{E.~Bertin}\inst{24,25} \and
{D.~Brooks}\inst{26} \and
{M.~Carrasco~Kind}\inst{27,28} \and
{F.~J.~Castander}\inst{2,3} \and
{R.~Cawthon}\inst{29} \and
{A.~Choi}\inst{4} \and
{C.~Conselice}\inst{30,31} \and
{M.~Costanzi}\inst{32,33,34} \and
{L.~N.~da Costa}\inst{8,35} \and
{M.~E.~S.~Pereira}\inst{36} \and
{H.~T.~Diehl}\inst{23} \and
{P.~Doel}\inst{26} \and
{A.~Drlica-Wagner}\inst{37,23,38} \and
{S.~Everett}\inst{39} \and
{A.~E.~Evrard}\inst{40,36} \and
{B.~Flaugher}\inst{23} \and
{J.~Frieman}\inst{23,38} \and
{J.~Garc\'ia-Bellido}\inst{10} \and
{D.~W.~Gerdes}\inst{40,36} \and
{D.~Gruen}\inst{41,42,43} \and
{R.~A.~Gruendl}\inst{27,28} \and
{J.~Gschwend}\inst{8,35} \and
{G.~Gutierrez}\inst{23} \and
{S.~R.~Hinton}\inst{44} \and
{D.~L.~Hollowood}\inst{39} \and
{K.~Honscheid}\inst{4,5} \and
{B.~Hoyle}\inst{45} \and
{D.~Huterer}\inst{36} \and
{D.~J.~James}\inst{46} \and
{K.~Kuehn}\inst{47,48} \and
{M.~Lima}\inst{49,8} \and
{M.~A.~G.~Maia}\inst{8,35} \and
{J.~L.~Marshall}\inst{50} \and
{F.~Menanteau}\inst{27,28} \and
{R.~Miquel}\inst{51,17} \and
{R.~Morgan}\inst{29} \and
{J.~Muir}\inst{42} \and
{R.~L.~C.~Ogando}\inst{8,35} \and
{A.~Palmese}\inst{23,38} \and
{F.~Paz-Chinch\'{o}n}\inst{27,52} \and
{W.~J.~Percival}\inst{53,54} \and
{A.~A.~Plazas~Malag\'on}\inst{55} \and
{M.~Rodriguez-Monroy}\inst{15} \and
{V.~Scarpine}\inst{23} \and
{M.~Schubnell}\inst{36} \and
{S.~Serrano}\inst{2,3} \and
{M.~Smith}\inst{56} \and
{M.~Soares-Santos}\inst{36} \and
{E.~Suchyta}\inst{57} \and
{M.~E.~C.~Swanson}\inst{27} \and
{G.~Tarle}\inst{36} \and
{D.~Thomas}\inst{58} \and
{C.~To}\inst{41,42,43} \and
{D.~L.~Tucker}\inst{23} \and
{T.~N.~Varga}\inst{59,60}\\ 
(DES Collaboration) 
}    

\institute{Institute of Theoretical Astrophysics, University of Oslo. P.O. Box 1029 Blindern, NO-0315 Oslo, Norway.  \and
Institut d'Estudis Espacials de Catalunya (IEEC), 08034 Barcelona, Spain. \and
Institute of Space Sciences (ICE, CSIC),  Campus UAB, Carrer de Can Magrans, s/n,  08193 Barcelona, Spain.  \and
Center for Cosmology and Astro-Particle Physics, The Ohio State University, Columbus, OH 43210, USA.  \and
Department of Physics, The Ohio State University, Columbus, OH 43210, USA.  \and
Max-Planck-Institut für Astrophysik, Karl-Schwarzschild Str. 1, 85741 Garching, Germany.  \and
Instituto de Astrofisica de Canarias, E-38205 La Laguna, Tenerife, Spain.  \and
Laborat\'orio Interinstitucional de e-Astronomia - LIneA, Rua Gal. Jos\'e Cristino 77, Rio de Janeiro, RJ - 20921-400, Brazil.  \and
Universidad de La Laguna, Dpto. Astrof\'isica, E-38206 La Laguna, Tenerife, Spain.  \and
Instituto de Fisica Teorica UAM/CSIC, Universidad Autonoma de Madrid, 28049 Madrid, Spain.  \and
Institute of Cosmology and Gravitation, University of Portsmouth, Dennis Sciama Building, Burnaby Road, Portsmouth PO1 3FX, United Kingdom.  \and
School of Physics and Astronomy, Sun Yat-sen University, 2 Daxue Road, Tangjia, Zhuhai, 519082, China.  \and
Instituto de F\'{i}sica Te\'orica, Universidade Estadual Paulista, S\~ao Paulo, Brazil.  \and
ICTP South American Institute for Fundamental Research.  \and 
Instituto de F\'{\i}sica Te\'orica, Universidade Estadual Paulista, S\~ao Paulo, Brazil.  \and
Centro de Investigaciones Energ\'eticas, Medioambientales y Tecnol\'ogicas (CIEMAT), Madrid, Spain.  \and
Port d'Informaci\'o Cient\'{i}fica (PIC), Campus UAB, C. Albareda s/n, 08193 Bellaterra (Barcelona), Spain.  \and
 Institut de F\'isica d'Altes Energies (IFAE), The Barcelona Institute of Science and Technology, Campus UAB, 08193 Bellaterra (Barcelona) Spain.  \and
Jet Propulsion Laboratory, California Institute of Technology, 4800 Oak Grove Dr., Pasadena, CA 91109, USA.  \and
Instituto de F\'isica Gleb Wataghin, Universidade Estadual de Campinas, 13083-859, Campinas, SP, Brazil.  \and
Department of Astronomy/Steward Observatory, University of Arizona, 933 North Cherry Avenue, Tucson, AZ 85721-0065, USA.  \and
Departamento de F\'{\i}sica Matem\'atica, Instituto de F\'isica, Universidade de S\~ao Paulo,CP 66318, S\~ao Paulo, SP, 05314-970, Brazil.  \and
Laborat\'orio Interinstitucional de e-Astronomia, Rua General Jos\'e Cristino, 77, S\~ao Crist\'ov\~ao, Rio de Janeiro, RJ, 20921-400, Brazil.  \and
Fermi National Accelerator Laboratory, P. O. Box 500, Batavia, IL 60510, USA.  \and
CNRS, UMR 7095, Institut d'Astrophysique de Paris, F-75014, Paris, France.  \and
Sorbonne Universit\'es, UPMC Univ Paris 06, UMR 7095, Institut d'Astrophysique de Paris, F-75014, Paris, France.  \and
Department of Physics \& Astronomy, University College London, Gower Street, London, WC1E 6BT, UK.  \and
Center for Astrophysical Surveys, National Center for Supercomputing Applications, 1205 West Clark St., Urbana, IL 61801, USA.  \and
Department of Astronomy, University of Illinois at Urbana-Champaign, 1002 W. Green Street, Urbana, IL 61801, USA.  \and
Physics Department, 2320 Chamberlin Hall, University of Wisconsin-Madison, 1150 University Avenue Madison, WI  53706-1390.  \\ \and
Jodrell Bank Center for Astrophysics, School of Physics and Astronomy, University of Manchester, Oxford Road, Manchester, M13 9PL, UK.  \and
University of Nottingham, School of Physics and Astronomy, Nottingham NG7 2RD, UK.  \and
Astronomy Unit, Department of Physics, University of Trieste, via Tiepolo 11, I-34131 Trieste, Italy.  \and
INAF-Osservatorio Astronomico di Trieste, via G. B. Tiepolo 11, I-34143 Trieste, Italy.  \and
Institute for Fundamental Physics of the Universe, Via Beirut 2, 34014 Trieste, Italy.  \and 
Observat\'orio Nacional, Rua Gal. Jos\'e Cristino 77, Rio de Janeiro, RJ - 20921-400, Brazil.  \and
Department of Physics, University of Michigan, Ann Arbor, MI 48109, USA.  \and
Department of Astronomy and Astrophysics, University of Chicago, Chicago, IL 60637, USA.  \and
Kavli Institute for Cosmological Physics, University of Chicago, Chicago, IL 60637, USA.  \and
Santa Cruz Institute for Particle Physics, Santa Cruz, CA 95064, USA.  \and
 Department of Astronomy, University of Michigan, Ann Arbor, MI 48109, USA.  \and
Department of Physics, Stanford University, 382 Via Pueblo Mall, Stanford, CA 94305, USA.  \and
 avli Institute for Particle Astrophysics \& Cosmology, P. O. Box 2450, Stanford University, Stanford, CA 94305, USA.  \and
SLAC National Accelerator Laboratory, Menlo Park, CA 94025, USA.  \and
School of Mathematics and Physics, University of Queensland,  Brisbane, QLD 4072, Australia.  \and
Faculty of Physics, Ludwig-Maximilians-Universit\"at, Scheinerstr. 1, 81679 Munich, Germany.  \and
Center for Astrophysics $\vert$ Harvard \& Smithsonian, 60 Garden Street, Cambridge, MA 02138, USA.  \and
Australian Astronomical Optics, Macquarie University, North Ryde, NSW 2113, Australia.  \and
Lowell Observatory, 1400 Mars Hill Rd, Flagstaff, AZ 86001, USA.  \and
Departamento de F\'isica Matem\'atica, Instituto de F\'isica, Universidade de S\~ao Paulo, CP 66318, S\~ao Paulo, SP, 05314-970, Brazil.  \and
George P. and Cynthia Woods Mitchell Institute for Fundamental Physics and Astronomy, and Department of Physics and Astronomy, Texas A\&M University, College Station, TX 77843,  USA.  \and
Instituci\'o Catalana de Recerca i Estudis Avan\c{c}ats, E-08010 Barcelona, Spain.  \and
Institute of Astronomy, University of Cambridge, Madingley Road, Cambridge CB3 0HA, UK.  \and
Department of Physics and Astronomy, University of Waterloo, 200 University Ave W, Waterloo, ON N2L 3G1, Canada.  \and
Perimeter Institute for Theoretical Physics, 31 Caroline St. North, Waterloo, ON N2L 2Y5, Canada.  \and
Department of Astrophysical Sciences, Princeton University, Peyton Hall, Princeton, NJ 08544, USA.  \and
School of Physics and Astronomy, University of Southampton,  Southampton, SO17 1BJ, UK.  \and
Computer Science and Mathematics Division, Oak Ridge National Laboratory, Oak Ridge, TN 37831.  \and
Institute of Cosmology and Gravitation, University of Portsmouth, Portsmouth, PO1 3FX, UK.  \and
Max Planck Institute for Extraterrestrial Physics, Giessenbachstrasse, 85748 Garching, Germany.  \and
Universit\"ats-Sternwarte, Fakult\"at f\"ur Physik, Ludwig-Maximilians Universit\"at M\"unchen, Scheinerstr. 1, 81679 M\"unchen, Germany.  
}

% Abstract of the paper
\abstract
{The calibration and validation of scientific analysis in simulations is a fundamental tool to ensure unbiased and robust results in observational cosmology. In particular, mock galaxy catalogs are a crucial resource to achieve these goals in the measurement of baryon acoustic oscillation (BAO) in the clustering of galaxies. Here we present a set of $1\,952$  galaxy mock catalogs designed to mimic the Dark Energy Survey (DES) Year 3 BAO sample over its full photometric redshift range $0.6  < z_{\rm photo} < 1.1$. The mocks are based upon 488 ICE-COLA fast $N$-body simulations of full-sky light cones and were created by populating halos with galaxies, using a hybrid halo occupation distribution - halo abundance matching model. This model has ten free parameters, which were determined, for the first time, using an automatic likelihood minimization procedure. We also introduced a novel technique to assign photometric redshift for simulated galaxies, following a two-dimensional probability distribution with VIMOS Public Extragalactic Redshift Survey (VIPERS) data. 
The calibration was designed to match the observed abundance of galaxies as a function of photometric redshift, the distribution of photometric redshift errors, and the clustering amplitude on scales smaller than those used for BAO measurements. 
An exhaustive analysis was done to ensure that the mocks reproduce the input properties. Finally, mocks were tested by comparing the angular correlation function $w(\theta)$, angular power spectrum $C_\ell$, and projected clustering $\xi_p(r_\perp)$ to theoretical predictions and data. The impact of volume replication in the estimate of the covariance is also investigated. The success in accurately reproducing  the photometric redshift uncertainties and the galaxy clustering as a function of redshift render this mock creation pipeline as a benchmark for future analyses of photometric galaxy surveys.
}

%%%%%%%%%%%%%%%%%%%%%%%%%%%%%%%%%%%%%%%%%%%%%%%%%%
\maketitle

%%%%%%%%%%%%%%%%% BODY OF PAPER %%%%%%%%%%%%%%%%%%
\section{Introduction} \label{sec:intro}

Over recent years, a large international effort has been focused on constraining the dark energy properties, measuring the cosmological parameters with high accuracy, and testing the Lambda cold dark matter (\LCDM) paradigm. This led to the development of new techniques and data combinations that allow tighter constraints: cosmic microwave background (CMB); Type Ia supernovae (SNe Ia); galaxy clustering (GC); weak lensing (WL); baryon acoustic oscillation (BAO); etc. Especially in recent years,  BAO \citep{Peebles1970,SZ1970} has become a powerful alternative to building the Hubble diagram, which now allows one  estimate cosmological parameters by itself \citep{Percival2007a,Beutler2011,eBOSS2020}.\\
Most of the probes mentioned before require the measurement of redshifts with high fidelity, giving more significance to spectroscopic surveys (e.g.,  the 
WFC3 Infrared Spectroscopic Parallel Survey - WISP\footnote{\url{http://wisps.ipac.caltech.edu}},
Baryon Oscillation Spectroscopic Survey - BOSS\footnote{\url{http://www.sdss3.org/surveys/boss.php}},
Euclid\footnote{\url{https://www.euclid-ec.org/}},
Wide-Field Infrared Survey Telescope - WFIRST\footnote{\url{https://wfirst.gsfc.nasa.gov/}},
Dark Energy Spectroscopic Instrument - DESI\footnote{\url{https://www.desi.lbl.gov/}}).
However, photometric surveys have some advantages over the spectroscopic ones. In particular, every observed galaxy can be used in the cosmological analysis although in practice is necessary to select a galaxy population that presents a prominent spectral feature that can be captured with broadband filters. Besides, many successful techniques are used to estimate true redshifts given observed photometric redshifts (\photoz) within a given uncertainty. 
Then, the statistical power makes imaging surveys almost as competitive as spectroscopic surveys in the measurement of galaxy clustering.
One example of these techniques is the Directional Neighbourhood Fitting \citep[\DNF,][]{DeVicente2016}, which stands out as one of the most  robust and accurate determinations of \photoz,
and is thus used in several photometric surveys \citep[see e.g.,][]{Drlica-Wagner2018,Sevilla-Noarbe2021,EuclidCollaboration2020}. 
The Dark Energy Survey (DES)\footnote{\url{https://www.darkenergysurvey.org}} is a badge example. DES has mapped the southern sky for six years covering an area of $\sim$ 5\,000 deg$^{2}$ and has recorded data from a few hundred million distant galaxies.  
These numbers will be pushed even beyond by future projects (e.g., the Legacy Survey of Space and Time - Rubin LSST\footnote{\url{https://www.lsst.org}}, Spectro-Photometer for the History of the Universe, Epoch of Reionization, and Ices Explorer - SPHEREx\footnote{\url{http://spherex.caltech.edu/}}).\\
For the large data sets that these projects produce, the calibration and replication of scientific analysis in simulations previous to \textit{unblinding} (procedure explained below), is a fundamental tool to ensure unbiased and robust results. This task requires the fulfillment of two requirements: $(i)$ a realistic simulation of the observed cosmological volume with a final galaxy catalog that mimics the data and $(ii)$ a large number of realizations varying the initial conditions that allows a full control of statistical uncertainties. The negative effects of introducing simulated volume replications to achieve the first requisite should not be underestimated. For example, the consequence of over-estimation of the covariances was found in the mocks used for this work. Accomplishing the second requirement by using pure $N$-body simulations is computationally impossible when the number of needed realizations is hundreds or thousands. Using approximate methods allows to have the desired number of runs with less computational resources \citep[see e.g.,][]{Coles91,Scoccimarro2002,Koda_cola,avila2015,Chuang2015,izard2018}. All these methods reduce the resolution of the simulation on small scales in exchange for computing speed. But when we focus our study on BAO scales, as the purpose of this work, it has been shown that the accuracy of these approximate methods is more than sufficient for a precise analysis \citep{Chuang2015b,Lippich2019,blot2020}. For example,  \cite{ice-cola} demonstrates that the ICE-COLA method yields a matter power spectrum within $1\%$ for $k\lesssim1~h~$Mpc$^{-1}$ and a halo mass function within $5\%$ of those in the $N$-body. Nowadays, all cosmological surveys need to develop their own galaxy mock catalogs in order to properly simulate the characteristics of the data. BOSS, eBOSS and the first year of DES data (DES Y1) for example have designed their own mocks \citep{Manera13,Chuang2015,Chuang2015b,Kitaura2016,Avila2018,Cheng2021}.
The BAO analysis using the first three years of DES data (DES Y3) is structured in three papers: \cite{y3baosample} presents a systematics analysis of the galaxy sample for DES Y3 BAO measurement, this work describes the simulations used in the analysis and the main DES Y3 BAO paper presents the angular distance constrains and cosmology in \cite{y3mainbao}. An analogous work was made for DES Y1: the DES Y1 \sample was presented in \cite{y1baosample}, a description of the mocks was shown in \cite{Avila2018} and \cite{mainBAOY1} as the main DES Y1 BAO paper including a $\sim$ $4\%$ precision $D_A$ measurement. In this case, the work was accompanied by several method papers \citealp{Chan:2018gtc,Ross2017,camacho2019}.\\
A common process in the analysis of new Surveys data release is to \textit{blind} the data in certain ways. The implementation of a rigorous process of \textit{unblinding} can reduce or eliminate confirmation bias. A strict blinding strategy has been applied to this work. The final set of mocks, presented here, were completed before computing $\alpha$ (BAO shift parameter) on the final data vector, and before plotting the angular two-point correlation function or clustering $C_\ell$ of the DES Y3 \sample. Only three \textit{pre-unblinding} values of the angular clustering on scales lower than one degree (unused for the BAO analysis) were provided to calibrate the clustering amplitude of the mocks. Once the mocks were done and the data passed through the rigorous process of \textit{unblinding} we could compare the clustering both in configuration and harmonic space of the mocks with the final \textit{post-unblinding} measurements of the data. We refer the interested reader to \cite{y3mainbao} for more details about the \textit{unblinding} process of the data. 

This paper is arranged as follows. In section~\ref{sec:redfata} we briefly describe the reference sample. Then, section~\ref{sec:halocat} describes the main features of the used dark matter halo catalogs. On the one hand, we present the fast simulation used to perform the mocks for the analysis, and on the other hand, our benchmark pure $N$-body simulation. In section~\ref{sec:Galcat} we detail step by step the process we followed to create our galaxy catalogs, which match the main properties of the data. One of the most important aspects of our pipeline is the automatic calibration, which is exhaustively detailed in section~\ref{sec:Calib}. After describing how the mocks are created, in section~\ref{sec:resul} we compare them with the real data and theoretical models in terms of covariance matrices and clustering measurements, both in angular configuration space and angular harmonic space. Additionally, in section~\ref{sec:rep} we investigate the effect of replicated structures in the mocks. Finally, we conclude in section~\ref{sec:conclu} with our summary and conclusions.\\

%%%%%%%%%%%%%%%%%%%%%%%%%%%%%%%%%%%%%%%%%%%%%%%%%%%%%%%%%%%%%%%%%%%%%%%%%%%%%%%%%%%%%%%%%%%%%%%%%%%%%%%%%%%%%%%%%%%%%%%%%%%%%%%%%%%%
%----------------------------------------------------------------------------------------------------------------------------------
%%%%%%%%%%%%%%%%%%%%%%%%%%%%%%%%%%%%%%%%%%%%%%%%%%%%%%%%%%%%%%%%%%%%%%%%%%%%%%%%%%%%%%%%%%%%%%%%%%%%%%%%%%%%%%%%%%%%%%%%%%%%%%%%%%%%

\section{Reference data} \label{sec:redfata}
The BAO analysis for the DES Y3 data set is based on the \gold catalog \citep{Sevilla-Noarbe2021}, which contains nearly $390$ million objects, with a depth reaching S/N $\sim10$ for extended objects up to $i<22.3 \ (AB)$, and top-of-the-atmosphere photometric accuracy under $3$ mmag. This data set has been compiled from the coaddition of nearly $40\,000$ exposures in the $grizY$ optical and near-infrared bands taken during the first three years of observations. It used the DECam instrument \citep{Flaugher:2015} from the Blanco Telescope in Cerro Tololo (Chile) and covers $5\,000$ deg$^{2}$ of the southern hemisphere.

The catalog includes positional, photometric, and morphological information, using a multi-epoch, multi-band fitting procedure of the object's shape in every exposure where the object is present (the Single Object Fitting, or \var{SOF}, method). This is the basis of all the measurements of the object mentioned before. In addition, the \gold catalog contains flagging information to assess the quality of the measured object, and ancillary survey information about low-quality regions in the sky, and survey properties in general (seeing, airmass, etc).

From the \gold, we select a sample of red galaxies used to measure the BAO scale, with a color selection similar to the DES Y1 analysis presented in \citet{y1baosample}. The DES Y3 selection shows an increase in the number density of galaxies (due to improvements on the DESDM\footnote{DES Data Management in National Center for Supercomputing Applications \citep[NCSA,][]{desdm}.} reduction), which allows us to extend the redshift range of the analysis to photometric redshift $z_{\rm photo}<1.1$ with $i<22.3 \ (AB)$. The selection applied to create the \sample is summarized in Table~\ref{tab:bao_selection}  \citep[details can be found in][]{y3baosample}.

\begin{table*}
\begin{center}
\caption{Selection process to create the BAO sample in DES Y3}
\label{tab:bao_selection}
\begin{tabular}{c c}
\hline
\hline
Keyword & Cut\\
\hline
Gold & observations present in the \gold catalog  \\
Quality & \verb|FLAGS_GOLD=0|  \\
Footprint & $4\,108.47$ deg$^{2}$  \\
Color selection & $(i-z) + 2.0(r-i)>1.7$  \\
Completeness cut & $i<22.3$  \\
Flux Selection & $17.5< i<19 + 3$ \, \ZMEAN  \\ 
Star-galaxy separation & \verb|EXTENDED_CLASS_MASH_SOF=3|  \\
\DNF \photoz range & $[0.6-1.1]$\\
\end{tabular}
\end{center}
\footnotesize
\captionsetup{width=.57\textwidth}
\captionsetup{font=small}
\caption*{We refer to \citet{Sevilla-Noarbe2021} and \citet{y3baosample} and online documentation for details about the meaning of the cuts.}
\end{table*}

\begin{figure}
    \begin{center}
        \includegraphics[width=0.5\textwidth]{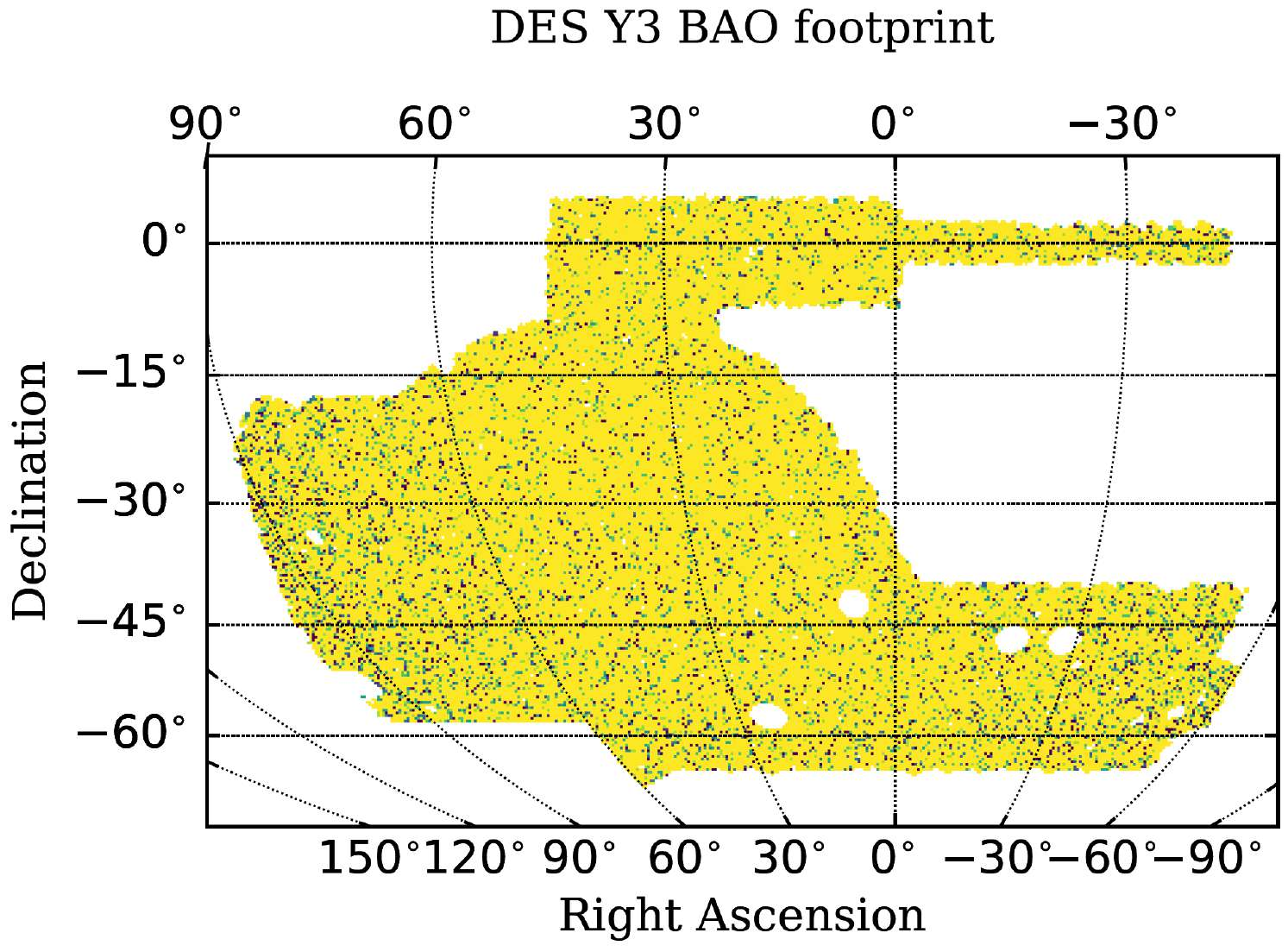}
    \end{center}
    \caption{DES Y3 \sample footprint, covering $4\,108.47$ deg$^{2}$ of the southern sky.}
    \label{fig:baofoot}
\end{figure}

Furthermore, the footprint mask is selected accordingly removing regions with depth less than 22.3, plus additional quality cuts explained in the aforementioned references. We use all HEALPix maps with \verb|NSIDE=4096| found in the online release\footnote{\url{https://www.darkenergysurvey.org/the-des-project/data-access/}}.
In Fig.~\ref{fig:baofoot}, we show the angular distribution of the BAO footprint, covering $4\,108.47$ deg$^{2}$

One of the most critical aspects of any photometric analysis is the measurement of the redshift. For the \sample, we characterize the $N(z)$ (the true redshift distribution) in each tomographic bin using the ``VIMOS Public Extragalactic Redshift Survey'' \citep[VIPERS]{vipers} catalog as a reference, since it is a complete sample from redshift above $z=0.5$ up to $i=22.5 \ (AB)$. VIPERS observed in two fields, named W1 and W4, both overlapping DES. The total overlap area is $16.324$ deg$^{2}$ This provides, after several selection processes described in \cite{y3baosample}, a final sample of $8\,362$ galaxies with spectroscopic redshift, $z_{\mbox{\tiny \rm  VIPERS}}$, available for redshift calibration. \cite{y3baosample} use VIPERS to validate the performance of the photometric redshift (called \ZMEAN in the DES catalogs) and to estimate the $true$ redshift (\ZMC) distribution of the DES Y3 \sample. \DNF predicts \ZMEAN as the best-value in the fitted hyper-plane and also defines \ZMC as the closest friend. In this work, we use these overlapping galaxies for the opposite purpose, assigning $z_{\rm photo}$ to the simulated galaxies. In other words, we re-sample the $z_{\rm spec}$ vs $z_{\rm photo}$ diagram found from VIPERS to assign $z_{\rm photo}$ to mock galaxies.\\

%%%%%%%%%%%%%%%%%%%%%%%%%%%%%%%%%%%%%%%%%%%%%%%%%%%%%%%%%%%%%%%%%%%%%%%%%%%%%%%%%%%%%%%%%%%%%%%%%%%%%%%%%%%%%%%%%%%%%%%%%%%%%%%%%%%%
%----------------------------------------------------------------------------------------------------------------------------------
%%%%%%%%%%%%%%%%%%%%%%%%%%%%%%%%%%%%%%%%%%%%%%%%%%%%%%%%%%%%%%%%%%%%%%%%%%%%%%%%%%%%%%%%%%%%%%%%%%%%%%%%%%%%%%%%%%%%%%%%%%%%%%%%%%%%
\section{Halo light cone catalogs} \label{sec:halocat} % used for referring to this section from elsewhere

In this section, we describe the halo catalogs from dark matter simulations used in this paper. We start by describing the ICE-COLA fast simulations and then our benchmark pure $N$-body simulation, MICE \textit{Grand Challenge}. It is important to make clear that both sets of simulations share the same cosmology, mass resolution, and halos found with a Friends of Friends algorithm.

\subsection{ICE-COLA fast simulations} \label{SUBsec:cola} % used for referring to this section from elsewhere
To build a large number of mocks we use a set of $488$  fast $N$-body simulations generated with the ICE-COLA code \citep{ice-cola}. The COmoving Lagrangian Acceleration (COLA) method solves for the evolution of the matter density field using second-order Lagrangian Perturbation Theory (2LPT) combined with a Particle-Mesh (PM) solver to integrate the particle orbits at small scales, where 2LPT start to deviate from the full $N$-body solution \citep{Tassev2013,cola}. The ICE-COLA code extends on this method to produce on-the-fly light cone halo catalogs and weak lensing maps \citep{izard2018}.

The simulations use $2\,048^3$ particles in a box of size of $1\,536$~Mpc~$h^{-1}$ to match the mass resolution of the MICE \textit{Grand Challenge} simulation (see Sect.~\ref{SUBsec:mice}).  Here we use the optimal code parameters found in \citet{ice-cola}, namely $40$ time-steps, a starting redshift of $z_{\rm ini}=19$ and a PM grid of $27$ times the number of particles. Halos are found with a Friends of Friends (FOF) algorithm with linking length $b=0.2$. We refer the interested readers to \citet{cola} and \citet{ice-cola} for more complex analysis and thorough validation of the method.\\
It is important to know the limitations of fast simulations to be able to use them in the range of scales where they agree with $N$-body, but also (moderately) beyond those scales with a systematic error that can be quantified (i.e, under control).

\subsection{MICE \textit{Grand Challenge} simulation} \label{SUBsec:mice} % used for referring to this section from elsewhere

\begin{figure}
    \begin{center}
        \includegraphics[width=0.46\textwidth]{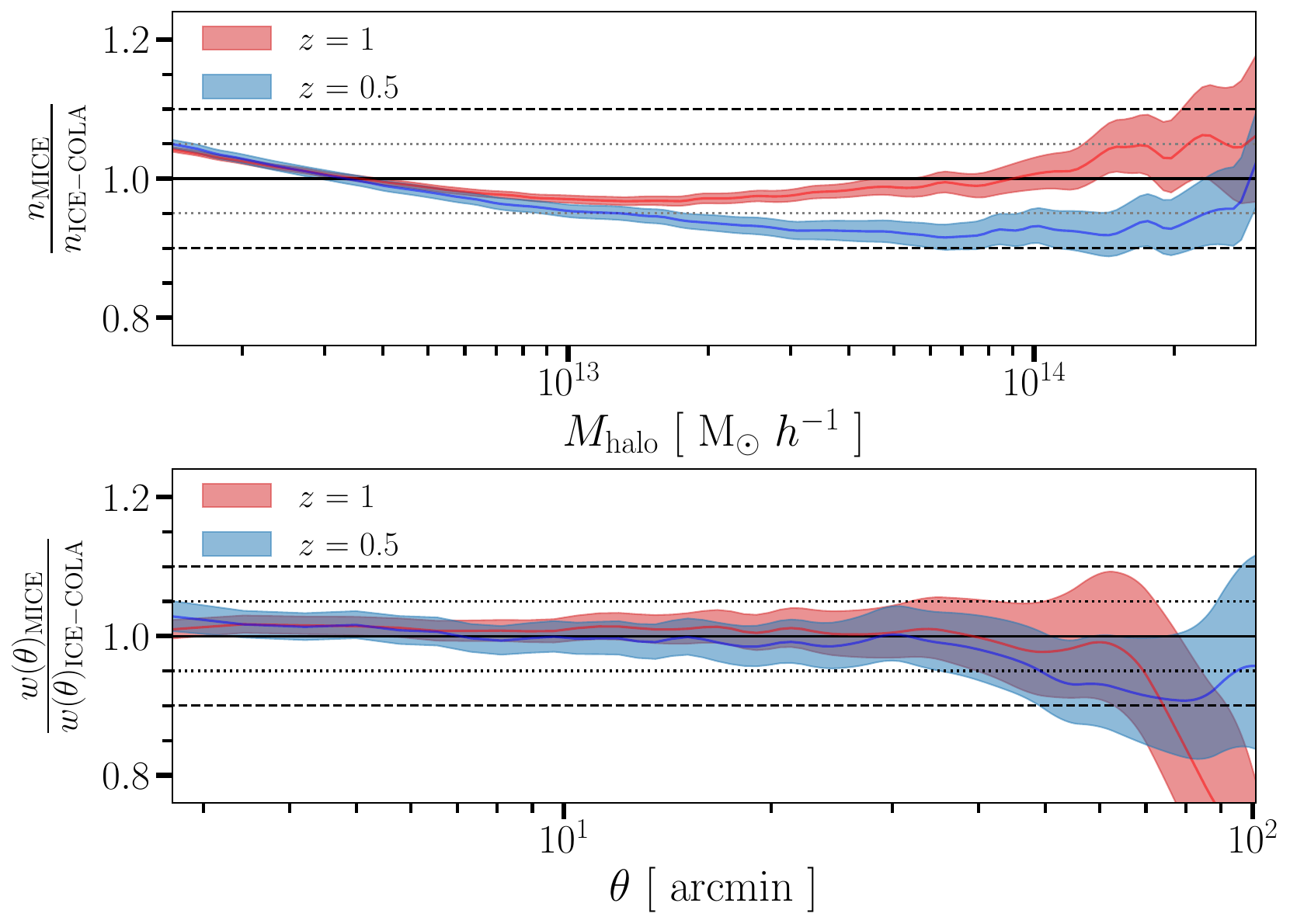}
    \end{center}
    \caption{\textit{TOP PANEL}: Ratio of the Halo Mass Function between MICE and ICE-COLA. Blue for redshift $z=0.5$ and red for $z=1$. Shaded areas correspond to the standar deviation of the $488$ ICE-COLA runs. \textit{BOTTOM PANEL}: Ratio between the MICE and the ICE-COLA  ACF for a sample of halos with $M_{\rm halo} > 1.46 \times 10^{12}$~M$_{\odot}$.}
    \label{fig:hmf_gc_micecola}
\end{figure}

As was mentioned before, it is really important to validate the range of scales where we can trust fast simulations. In our case, we use the MICE \textit{Grand Challenge}\footnote{More information is available at \url{http://maia.ice.cat/mice/}} \citep[][MICE hereafter]{MICE3,MICE2,MICE1}, simulation as the benchmark $N$-body run. MICE is an all-sky light cone $N$-body simulation evolving $4\,096^3$ dark-matter particles in a $\sim 29$~Gpc$^{3}$~$h^{-3}$ comoving volume. The assumed cosmology corresponds to the best-fit of WMAP five-year data \citep{Dunkley09}. This is consistent with a flat \LCDM model with $\Omegam=0.25$, $\OmegaL=0.75$, $\Omegab=0.044$, $n_s=0.95$, $\sigma_8=0.8$ and $h=0.7$.\\
As our reference simulation, its cosmological parameters are also used for our ICE-COLA runs, where only the initial condition changes among the $488$ fast simulations. An exhaustive validation of the simulations used here has been done in \citet{ice-cola}, finding a matter power spectrum within 1$\%$ for $k\lesssim$1~$h$~Mpc$^{-1}$ and demonstrating that ICE-COLA fast simulation can perfectly be used for BAO purposes. In Fig.~\ref{fig:hmf_gc_micecola}, we compare the halo masses and the clustering for redshift $z=0.5$ (blue) and $z=1$ (red). The top panel shows the ratio between the MICE and the ICE-COLA halo mass function, the lowest halo mass plotted here correspond to the mass limit of the ICE-COLA mock used in this paper, $1.46 \times 10^{12}$~M$_{\odot}$ (50 particles). Both simulations are consistent, as was found in \citet{ice-cola}, within an accuracy of $\sim$ $5\%$. The bottom panel of Fig.~\ref{fig:hmf_gc_micecola} shows the ratio between the MICE and the ICE-COLA angular two-point correlation function (ACF). In this case, the clustering is calculated using halos with more than 50 particles in a full-octant comoving output shell of width $125$ Mpc and $166$ Mpc for redshift $z=0.5$ and  $z=1$, respectively. This threshold of $50$ particles is not set deliberately but is a resulting minimum number of particles of the halos of the mocks presented in this work.
For clustering, the accuracy is within $\sim5\%$ up to scales of one degree. For higher angular distances the error increase, especially for redshift $z=1$ where these scales correspond to larger 3D distances.\\

\section{Galaxy light cone catalogs} \label{sec:Galcat} % used for referring to this section from elsewhere

With the simulations and the corresponding ICE-COLA halo catalogs presented in the previous section, we can start now by describing, step by step, the mechanism used to construct a galaxy mock beginning with a halo catalog. Some of our recipes described below closely follows \cite{Carretero2015} and we use a similar hybrid Halo Occupation Distribution - Halo Abundance Matching model modeling strategy presented by \cite{Avila2018} in the analysis of the DES Y1 data release.
Before going into the details, it is important to remark that simulated box replications are needed to have light cones reaching higher redshifts than $0.6$ (corresponding limiting redshift if we set the light cone origin at the center of a box-size of $1\,536$ Mpc $h^{-1}$) and covering the DES Y3 footprint. Four boxes on each Cartesian direction are needed (a total of $64$ simulated boxes) to create a full-sky light cone up to redshift $\sim 1.4$. The implications that these replications have on the analysis are discussed in more detail in Sect.~\ref{sec:rep}.\\
 Here, we create a mock of galaxies for BAO analysis from a halo catalog of a fast simulation. However, the described procedure can be applied to any sort of halo catalog to mimic any kind of galaxies samples. A key aspect that makes this pipeline successful is the inclusion of an automatic calibration, as discussed in Sect.~\ref{sec:Calib}.\\

\subsection{Halo occupation distribution} \label{SUBsec:HOD} % used for referring to this section from elsewhere
\begin{figure}
    \begin{center}
        \includegraphics[width=0.47 \textwidth]{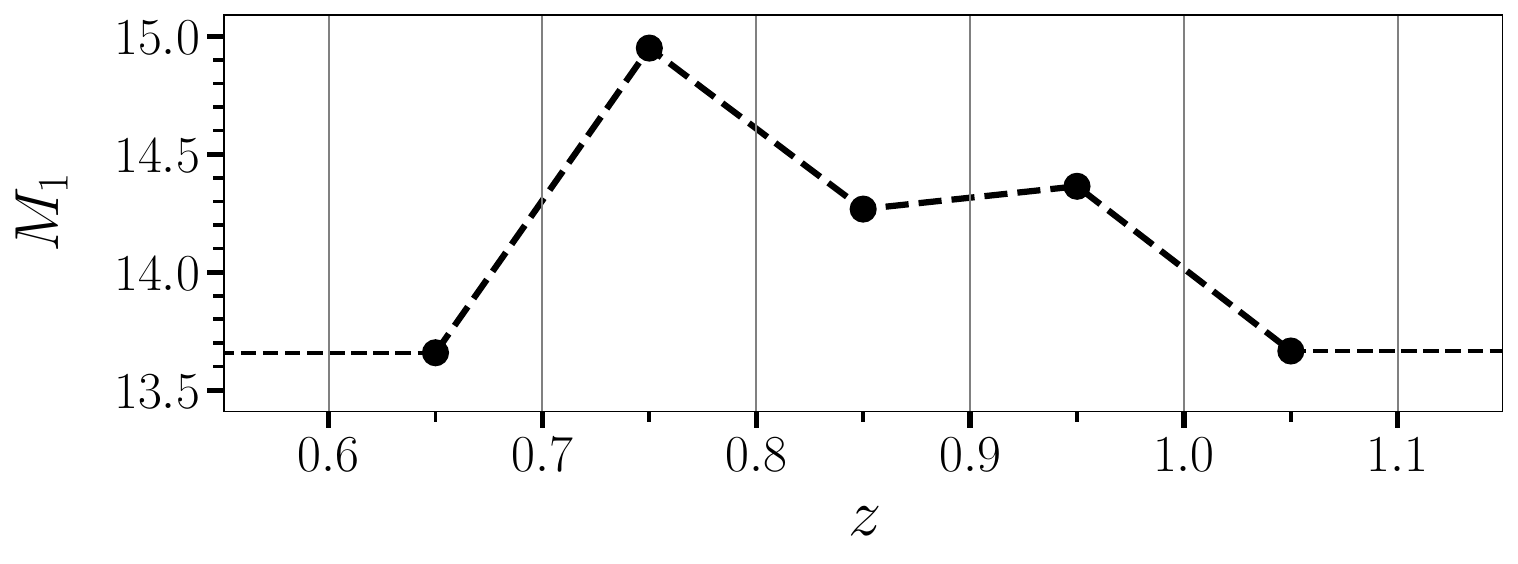}
    \end{center}
    \caption{Evolution of the first HOD free parameter $M_{1}$. One value for each tomographic bin. Dashed line correspond to the interpolation assumed for all the redshift range.}.
    \label{fig:Mi_evol}
\end{figure}

The relation between galaxies and halos is not univocal, as one halo can harbor more than one galaxy. Furthermore, the more massive the halo the higher the number of galaxies it has, reaching quantities of hundreds of galaxies in a single halo. The halo occupation distribution \citep[HOD;][]{Jing1998,Benson2000,Seljak2000}, describes the relation between halos and galaxies, in terms of several parameters. In other words, the HOD tells us how many galaxies a halo of a given mass has on average, $\langle N|M_{\rm halo} \rangle $.\\
Two different functions are needed to describe the HOD of a sample of galaxies. One for the central galaxies and another for satellites as they model clustering on different scales in the halo model. Centrals shape large scales (halo-halo correlations) and satellite small scales (intra-halo correlations). The complexity of the function can be as high as desired, in order to match the behavior of the sample with higher accuracy. We focus here on the large-scale structure, leaving aside a complex function that would allow us to model the small scales. Therefore, we assign to each halo, one central galaxy

\begin{equation}
    N_{\rm cent} = 1 , 
\label{eq:hodcen}
\end{equation}    
and a number of satellite galaxies following a Poisson distribution with mean

\begin{equation}
    N_{\rm sat} = \frac{M_{\rm halo}}{M_1} , 
\label{eq:hodsat}
\end{equation}
where $M_{\rm halo}$ is the mass of the halo, and $M_1$ is a free HOD parameter. This simple HOD is used to populate all halos with galaxies in the light cone. However, the particular \sample selection is a sub sample of this generic HOD assignment. Therefore, the final values of $N_{\rm cent}(M_{\rm halo})$ and $N_{\rm sat}(M_{\rm halo})$ that compose the \sample of the mocks differs from the expression defined on Eq.~\eqref{eq:hodcen} and \ref{eq:hodsat}.  \\
Once $N_{\rm cent}$ and $N_{\rm sat}$ values are determined, the next step is to populate the halos with galaxies following these HOD quantities. One central galaxy is placed in the center of each halo and the velocity is assumed to be equal to its host halo. On the other hand, satellites galaxies are distributed inside the halo following a spherical NFW \citep{NFW1996} profile. Concentrations, needed to model the NFW density profile, are taken from \cite{cosh2002} where the inputs are the mass and redshift of halos.
We also model the velocity of galaxies with simplistic assumptions, using a simple Gaussian distribution centered at the velocity of the host halo and assuming a standard deviation proportional to the velocity dispersion of it \citep{shdia2001,Carretero2015} 
\\
This first free parameter, $M_1$, allows to control the clustering: by increasing the parameter, we undersample the most massive halos decreasing the linear bias. It also introduces a 1-halo term that fades away as we increase $M_1$. More details can be found in \cite{Avila2018}. Fig.~\ref{fig:Mi_evol} shows the evolution of $M_1$ over the five tomographic bins where we assume a linear interpolation among these values. In Sect.~\ref{sec:Calib} is explained in detail how these values are obtained.\\

At this point, we already have a general galaxy catalog with positions and velocities, made by populating all halos on the light cone. The next step is to introduce a second free parameter by setting pseudo luminosities. This second HOD parameter allows setting a \sample by selecting only high luminosity galaxies from the general catalog.

\subsection{Pseudo-luminosity assignment} \label{SUBsec:Luminosity} % used for referring to this section from elsewhere
The best tracers for BAO signal are brightest galaxies \citep[see e.g.,][]{compa2013} and they represent a few percent of the total number of galaxies. Not all galaxies resulting from the previous step \ref{SUBsec:HOD} will enter into our selection to perform the BAO analysis. Therefore, this step is needed to select that few percent of galaxies.\\
An efficient way to select the tracers of our mocks is by assigning a pseudo luminosity $l_{\rm p}$ to all galaxies and then selecting the most luminous ones. To set the luminosities we rely on the halo abundance matching (HAM) techniques \citep{krav2004, conroy2006,guo2010}, where it is assumed that the most massive (luminous) galaxy lives in the most massive halo, the second most massive galaxy lives in the second most massive halo, and so on. On top of the mean (deterministic) relation between mass and luminosity assumed in the AM technique, we add some scattering to make this matching closer to observations. We model $l_{\rm p}$ with a Gaussian scatter around the halo mass $M_{\rm halo}$ in logarithmic scales:

\begin{equation}
    {\rm log} (l_{\rm p}) = {\rm log} (M_{\rm halo}) + \Delta_{\rm LM} . R^{gauss}_{\mu=0, \sigma=1}  ,
\label{eq:lumin}
\end{equation}
where $\Delta_{\rm LM}$ is our second free parameter which controls the amount of scatter. We note that $l_{\rm p}$ is modeled in arbitrary scales. The purposes of defining a luminosity for galaxies are two:
\begin{enumerate}
    \item It allows one to match the abundance and redshift distribution of data by selecting the most luminous galaxies. More details on Sect.~\ref{SUBsec:PhotoZ}.
    \item Its definition, and therefore the introduction of the second free parameter $\Delta_{\rm LM}$, also influences the clustering. As we decrease this value, lower mass halos go out of our selection and higher mass halos enter it, effectively increasing the bias.
\end{enumerate}

\begin{figure}
    \begin{center}
        \includegraphics[width=0.47
        \textwidth]{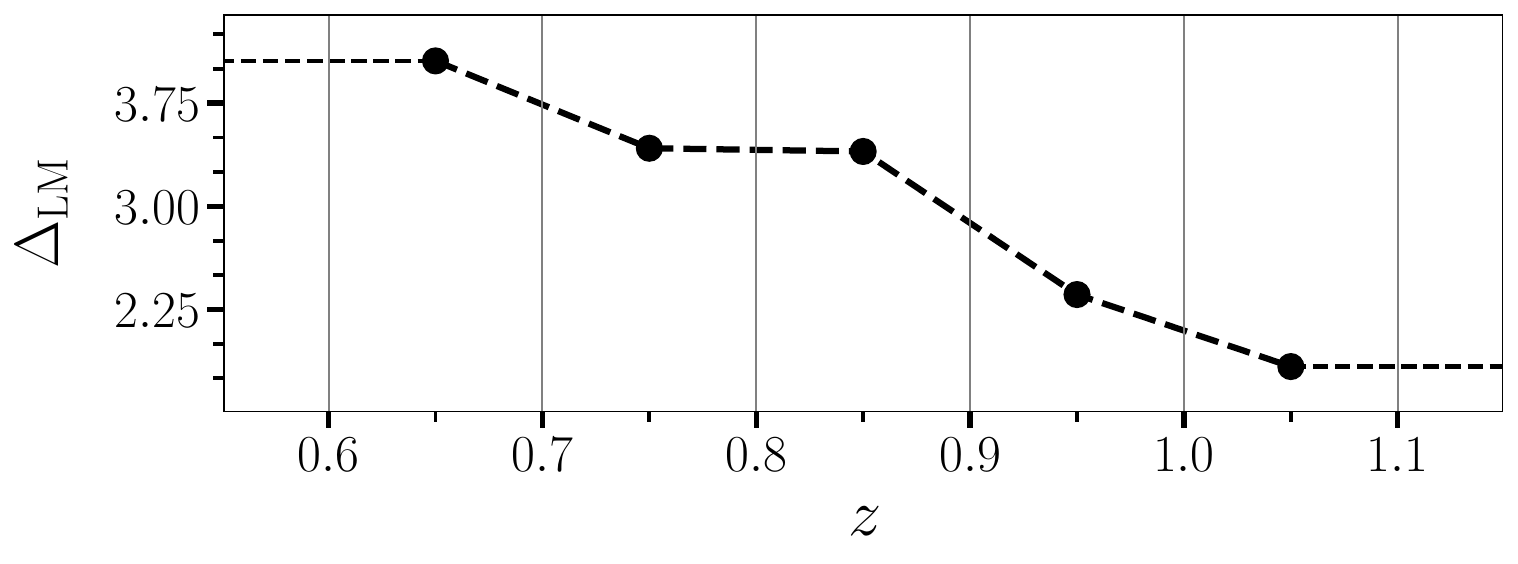}
    \end{center}
    \caption{Evolution of the second HOD parameter $\Delta_{\rm LM}$. One value for each tomographic bin. Dashed line corresponds to the interpolation assumed for all redshift range.}
    \label{fig:DeltaLM}
\end{figure}

Fig.~\ref{fig:DeltaLM} shows the evolution of this scatter parameter $\Delta_{\rm LM}$ as a function of redshift. In the same way as $M_{1}$, we assume for $\Delta_{\rm LM}$ one value for each tomographic bin and applying linear interpolation among these values as a function of redshift. The modeling explained in this subsection follows the same procedure used in \citet{Avila2018}. As was mentioned earlier, in Sect.~\ref{sec:Calib} it is explained how these values are obtained.

\subsection{Photometric redshifts} \label{SUBsec:PhotoZ} % used for referring to this section from elsewhere
\begin{figure}
    \begin{center}
        \includegraphics[width=0.47 \textwidth]{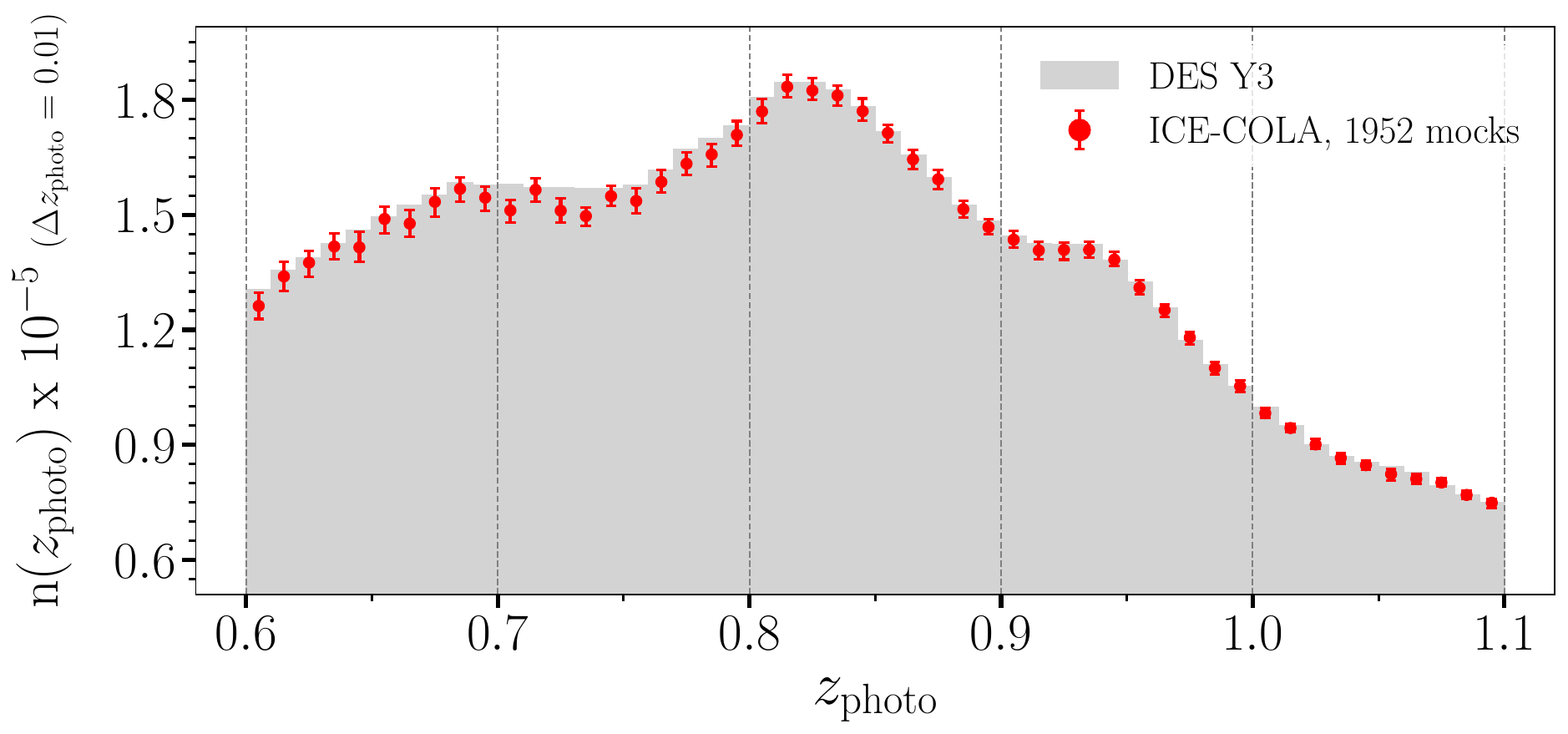}
    \end{center}
    \caption{Photometric redshift distribution of data and mocks. Gray histogram corresponds to DES Y3 \sample. Red points represent the average over the $1\,952$  ICE-COLA mocks and error bars denote the maximum and minimum.}
    \label{fig:PhotoZ_distribution}
\end{figure}

\begin{figure}
    \begin{center}
        \includegraphics[width=0.47
        \textwidth]{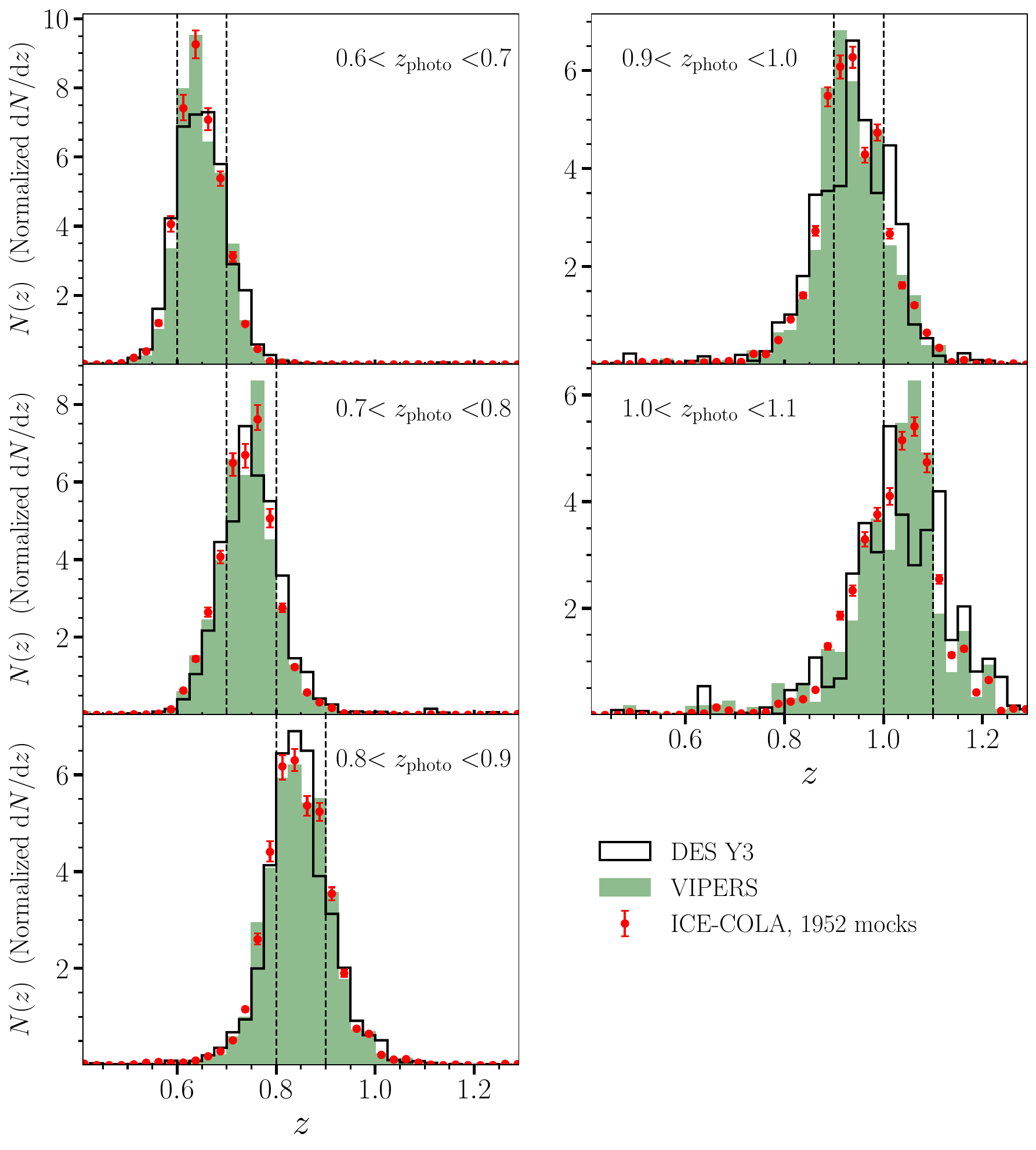}
    \end{center}
    \caption{True redshift $z_{\rm spec}$ distribution in each tomographic bin. Green filled histograms correspond to VIPERS, black lines represent the distribution for Y3 data and red point are the average over the $1\,952$  ICE-COLA mocks, while error bars correspond to the maximum and minimum. Histograms are normalized to have an integral of unity.}
    \label{fig:n_z}
\end{figure}

\begin{figure}
    \begin{center}
        \includegraphics[width=0.48 \textwidth]{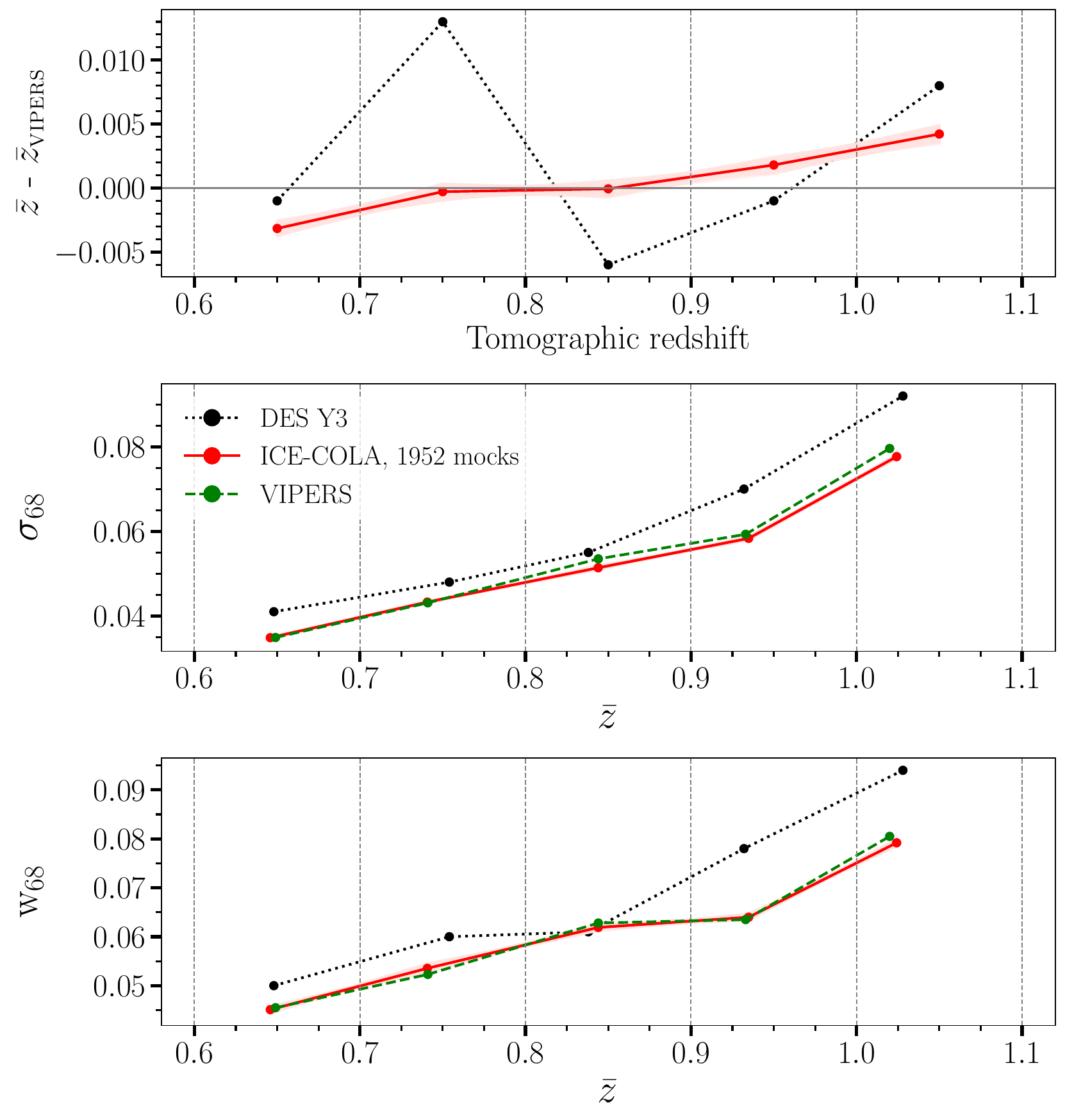}
    \end{center}
    \caption{Basic metrics for the \photoz validation on each tomographic bin. From
    top to bottom: difference between $\bar{z}$ and $\bar{z}_{\mbox{\tiny \rm  VIPERS}}$, evolution of $\sigma_{68}$ and $W68$ as a function $\bar{z}$. Black and red curves correspond to DES Y3 and ICE-COLA, respectively, and green lines on bottom panels represent VIPERS.}
    \label{fig:accuracy_n_z}
\end{figure}

This is perhaps one of the most dedicated and challenging steps of the mock creation pipeline. For the ICE-COLA mocks we know the true redshift and we need to model the \textit{observed} one for each simulated galaxy, contrary to what happens for observations. DES is a photometric survey, therefore the measurement of the redshift has a precision much lower than spectroscopic surveys. For example, \cite{y3baosample} show that the dispersion on the \photoz for the DES Y3 \sample is $\sigma_{68}=0.054$ on average for the five tomographic bins. $\sigma_{68}$ is defined as the value such as 68 per cent of the galaxies have $\lvert z_{\rm photo}-z_{\rm spec} \rvert / (1+z_{\rm spec}) < \sigma_{68}$. These uncertainties must be modeled in the simulation to have consistent clustering measurements.\\
Each galaxy in DES has an \textit{observed} photometric redshift \ZMEAN, derived from the magnitude measured in each filter. In addition, as was explained in Sect.\ref{sec:redfata}, there is a small sample of $8\,362$ galaxies for which we also have the true spectroscopic redshifts from VIPERS ($z_{\mbox{\tiny \rm  VIPERS}}$). The combination of using DES and VIPERS result in our mocks matching the abundance of DES Y3 BAO galaxies $n(\ZMEAN)$ and the redshift distribution $N(z_{\mbox{\tiny \rm  VIPERS}})$ on each tomographic bin of these galaxies present in both surveys.\\
We start by dividing the interval ~$\ZMEAN = [0.6,1.1]$ into $L$ thin bins of with $\Delta \ZMEAN = 0.01$. Then, according to the data, we can express the number of galaxies each $l$ bin has as $n(\ZMEAN^{l})$. This is the first condition we want to accomplish with the mocks: match the abundance of each $l$ thin photometric bin.\\
Secondly, we select $M$ spectroscopic bins of width $\Delta z_{\mbox{\tiny \rm  VIPERS}} = 0.025$, here bins are thicker because of the smaller number of VIPERS galaxies. Then, we can determine the probability of having a galaxy in a given pair of bins ($l$,$m$) as $P (\ZMEAN^{l},z_{\mbox{\tiny \rm  VIPERS}}^{m})$. It is important to remark that this matrix is built only using those DES galaxies which have a $z_{\mbox{\tiny \rm  VIPERS}}$. Mocks need to satisfy this 2D probability distribution $P (\ZMEAN,z_{\mbox{\tiny \rm  VIPERS}})$ and, at the same time, match the abundance of galaxies $n(\ZMEAN)$. By combining both, the number of galaxies an ICE-COLA mock should have at a given pair of bins($l$,$m$) can be calculated as

\begin{equation}
    A_{l,m} =  n(\ZMEAN^{l}) \times P (\ZMEAN^{l},z_{\mbox{\tiny \rm  VIPERS}}^{m}).\\
\label{eq:nice}
\end{equation}
The assignment of photometric redshifts $z_{\rm photo}$ to galaxies in mocks is then performed in two steps. Firstly, we separate the simulated galaxies into $L$ and $M$ bins and assigning $z_{\rm photo}$ by following the distribution $P (\ZMEAN,z_{\mbox{\tiny \rm  VIPERS}})$. And finally, we choosing from each ($l$,$m$) pair of bins the  $A_{l,m}$ most luminous galaxies, given the luminosities defined on Eq.~\eqref{eq:lumin}. The Fig.~\ref{fig:PhotoZ_distribution} shows the resulting $n(z_{\rm photo})$ for the mocks compared with data. Gray histograms correspond to DES Y3 \sample and red points with error bars represent the mocks. The agreement is almost perfect, as expected given that it is done by construction. On the other hand, to achieve a good match on redshift distribution $N(z_{\rm spec})$ for each tomographic bin is normally not so easy, but with this technique, it is also achieved by construction. This is shown in Fig.~\ref{fig:n_z} where filled green histograms correspond to VIPERS data while the black line denotes distribution for DES Y3. As in Fig.~\ref{fig:PhotoZ_distribution} red points represent the average of the mocks and error bars correspond to the maximum and minimum. The goal here was to assign $z_{\rm photo}$ in such a way that it gives a redshift distribution $N(z_{\rm spec})$ on each tomographic bin matching those of the DES Y3 BAO galaxies present in VIPERS.\\
Some important quantities must be compared for a correct \photoz validation, and are evaluated on each tomographic bin. These are the mean redshift ($\bar{z}$), the width of the $N(z)$ ($W68$, opposed to the dispersion), and the dispersion on the \photoz ($\sigma_{68}$). The performance of these quantities in the mocks are analyzed against VIPERS galaxies (for those we have both \ZMEAN and $z_{\mbox{\tiny \rm  VIPERS}}$) in Fig.~\ref{fig:accuracy_n_z}. The top panel shows the difference in $\bar{z}$ between the estimated true redshift for DES Y3 \sample (\ZMC, dotted black) and ICE-COLA mocks (red) against VIPERS ($z_{\mbox{\tiny \rm  VIPERS}}$). In the medium and bottom panel, we present the evolution of $\sigma_{68}$ and $W68$ as a function of $\bar{z}$ for each sample, respectively.
In all three quantities studied here, the agreement between VIPERS and the mocks is very satisfactory, showing a difference within $1\%$. Of course, our method to assign $z_{\rm photo}$ by construction should show a perfect match, with zero uncertainties. However, the small differences we see in Fig.~\ref{fig:accuracy_n_z} come from the fact that the sample of VIPERS galaxies ($8\,362$) is not fully representative of the 2D space $z_{\rm photo}$-$z_{\rm spec}$. We want to stress that the difference between $z_{\mbox{\tiny \rm  VIPERS}}$ and \ZMC represents an estimation of the uncertainty we have on the redshifts, and hence the precision of the mocks is well below those uncertainties. We refer the interested reader to \cite{y3baosample} for details on the performance of using VIPERS as a training sample for \photoz validation and $true$ redshift assignment of the \sample.\\
This step is different from what was done for DES Y1 by \citet{Avila2018}. Here we use an ``exact'' method which may propagate noise while \citet{Avila2018} used an ``analytical'' procedure (fitting a \textit{double skewed} Gaussian) which may miss some \photoz features.\\

\subsection{Masking}
\begin{figure}
    \begin{center}
    \includegraphics[width=0.47\textwidth]{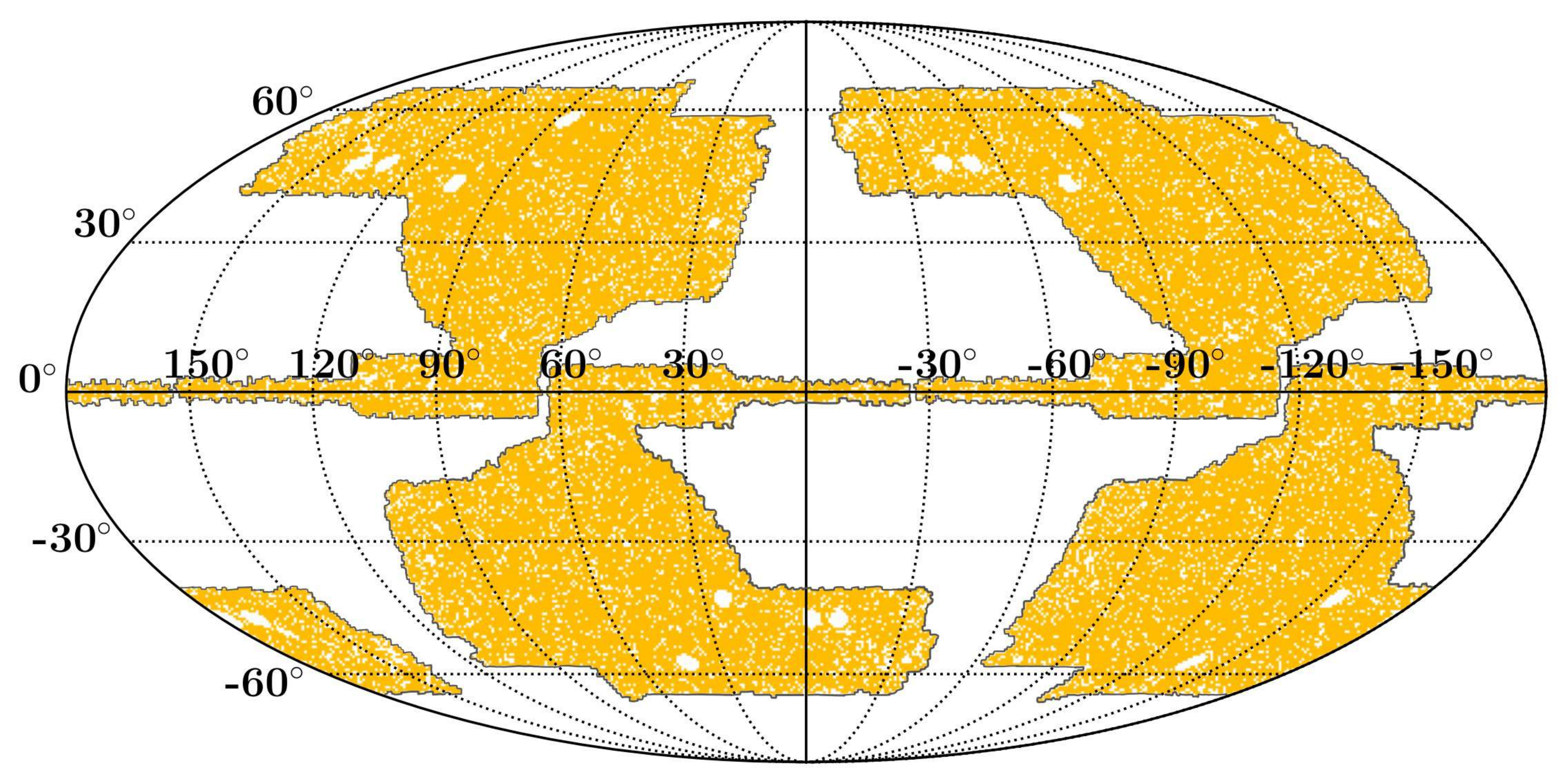}
    \cprotect\caption{Configuration of four DES Y3 BAO footprints on a full-sky. Mask are constructed with HEALPix assuming a resolution \verb|NSIDE=4096|.}
    \label{fig:Masking}
    \end{center}
\end{figure}

 label{SUBsec:masking} % used for referring to this section from elsewhere
Finally, our last step is to create galaxy mocks with the same footprint as the data. The angular mask of the DES Y3 \sample, described in Sect~\ref{sec:redfata}, has an area of $4\,108.47$ deg$^{2}$ and final ICE-COLA mocks must have the same characteristics with the same HEALPix resolution \verb|NSIDE=4096| as the data mask. To satisfy this and at the same time be efficient in creating as many catalogs as possible, four masks are placed on each full-sky light cone. This allows us to go from having $488$ ICE-COLA runs to having $1\,952$  BAO galaxy mocks at the end, quadrupling the number of simulations. This is illustrated in Fig.~\ref{fig:Masking} where four DES Y3 BAO footprints are placed, without overlapping, in a full-sky light cone. Although this configuration of masks maximizes the number of mocks, on Sect.~\ref{sec:rep} we return to this and analyze the negative implications it has on the covariance matrix.

%%%%%%%%%%%%%%%%%%%%%%%%%%%%%%%%%%%%%%%%%%%%%%%%%%%%%%%%%%%%%%%%%%%%%%%%%%%%%%%%%%%%%%%%%%%%%%%%%%%%%%%%%%%%%%%%%%%%%%%%%%%%%%%%%%%%
%----------------------------------------------------------------------------------------------------------------------------------
%%%%%%%%%%%%%%%%%%%%%%%%%%%%%%%%%%%%%%%%%%%%%%%%%%%%%%%%%%%%%%%%%%%%%%%%%%%%%%%%%%%%%%%%%%%%%%%%%%%%%%%%%%%%%%%%%%%%%%%%%%%%%%%%%%%%
\section{Calibration} \label{sec:Calib}
The pipeline used to generate the BAO mocks contains two free parameters per tomographic bin, $M_1$ and $\Delta_{\rm LM}$, as detailed in the previous section. This amounts to a total of ten free parameters that should be varied altogether to minimize the difference between the measurements and the mocks. In more detail, we want to minimize the discrepancy between the measured $w(\theta)$ and the one obtained from the mocks. Let us emphasize that ``$w(\theta)$'' in this section refers to the measurement of the angular correlation function calculated only for the three \textit{pre-unblinding} angular apertures $\theta=[0.58, 0.75,0.92]$. The main problem that we have to address is that we cannot obtain $w(\theta)$ from the mocks without first specifying the values of these parameters, generating the mocks, and measuring the observable quantity. If the parameter space were small we could attempt to vary the parameters one at a time, run a few cases, and try to determine approximately the best values for the parameters. However, a 10-dimensional parameter space and $\sim 500$ CPU hours to generate the mocks make it effectively impossible to follow this brute force approach.

In this work, we make use of the novel technique presented in \citet{calibration}, where an automatic calibration procedure is implemented into the pipeline to enable us to sample the parameter space and provide the values giving the best agreement with the data in a fully automatized way. We present the basic idea of this method, while we refer the reader to \citet{calibration} for all the details.

\begin{figure}
    \begin{center}
        \includegraphics[width=0.49\textwidth]{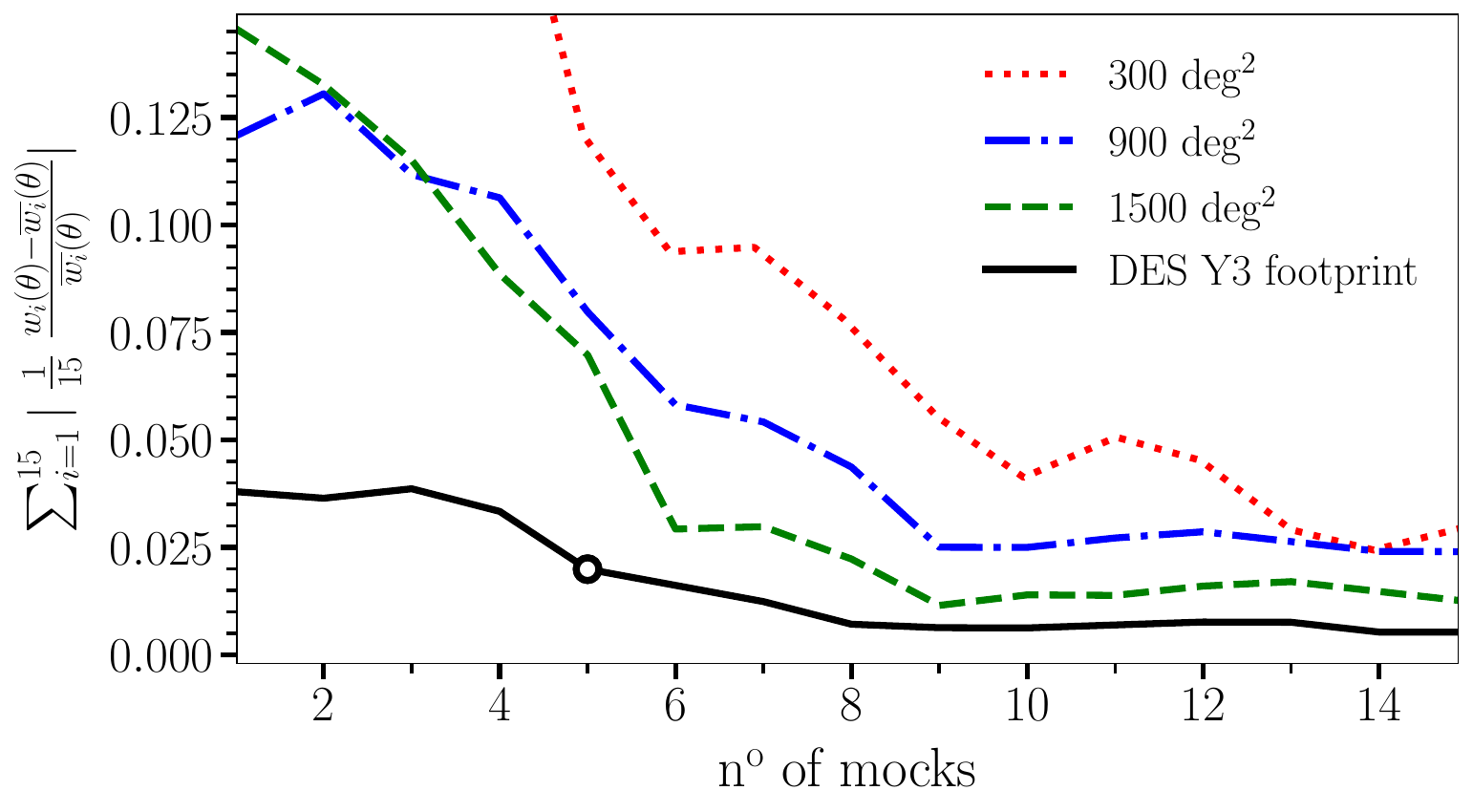}
    \caption{Agreement between $w(\theta)$ of a subset of mocks and the mean of $w(\theta)$ for all mocks as a function of the number of mocks used for the selection. The dotted red line stands for mocks of $300$ deg$^{2}$, while the dot-dashed blue line represents mocks of $900$ deg$^{2}$ and the dashed green one corresponds to mocks of $1\,500$ deg$^{2}$ The solid black line stands for mocks of $4\,108.47$ deg$^{2}$ with the mask of DES Y3 BAO data. The black open circle denotes the selection of the number of mocks and area used for the calibration in this analysis. The number $15$ on the y-axis corresponds to the number of clustering measurements averaged (three angular apertures times five redshift bins)}
    \label{fig:area_number_mocks}
    \end{center}
\end{figure}

The first step of the calibration is the determination of the minimal number of mocks that need to be used to get a statistically representative measurement of $w(\theta)$. This will allow us to calibrate with a subset of the mocks and use the best-fit parameters for all of them. In addition to the number of mocks, we also need to find the optimal area for calibration. If the area considered is too small, the determination of $w(\theta)$ will be affected by cosmic variance and it might not be statistically representative. To find the minimum number of mocks with the smallest area that we need to use for the calibration (based on a maximum feasible computation time), we have compared $w(\theta)$ from the selected mocks to the mean $\overline{w}(\theta)$ of the full set of mocks as a function of the number of mocks for different areas. The results are shown in Fig.~\ref{fig:area_number_mocks}. As can be seen in the figure, we need a large number of mocks of $300$ deg$^{2}$ (dotted red line) to obtain a $w(\theta)$ representative of the full sample. However, if we consider mocks with the same angular coverage as DES Y3 BAO data we can obtain a good representative of the full $w(\theta)$ by just using five mocks. This is the area and number of mocks used for the calibration and it is represented with a black empty circle in Fig.~\ref{fig:area_number_mocks}. We choose the combination number of mock-area requiring less computational time to get uncertainties within $2\%$. Note that this optimization of the area and number of mocks have been performed using a fixed value for the calibration parameters, $M_1=[13.5, 13.9, 14.5, 13.8, 13.2]$ and $\Delta_{\rm LM}=[1.06, 1.23,  1.97,  3.14,  2.21]$, but our goal here was to determine the size of the representative subset of mocks, not the agreement with the data yet. Therefore, these fixed values do not have a significant impact on the subset of mocks that will be used for the calibration. Moreover, we note that the chosen $2\%$ accuracy is somewhat arbitrary. As it can be seen in Fig.~\ref{fig:area_number_mocks}, considering more mocks provides a better agreement. However, we have verified that $2\%$ is enough for our purposes, guaranteeing
 uncertainties within $1\sigma$, and it still allows us to use a reduced number of mocks per point in the calibration.

Once we have determined how many mocks and which area we will use for the calibration, we need to start sampling the parameter space to determine the best-fit parameters. The main idea is to sample a given hypercube in the parameter space. In each point we generate five mocks using the DES Y3 BAO mask, measure $w(\theta)$, and compute the value of the $\chi^2$ of the measured $w(\theta)$ in the mocks to the real measurements:

\begin{equation}
    \chi^2=(w(\theta)_{\rm data}-w(\theta)_{\rm mocks})^{\rm T}C^{-1}(w(\theta)_{\rm data}-w(\theta)_{\rm mocks})\,.
\end{equation}\\
We note that $C$, which enters into the $\chi^2$, is the standard covariance matrix of the $1\,952$ mocks and takes into account the correlations between the different tomographic bins. To obtain $C$ we calculated $w(\theta)$ for the $1\,952$ mocks previously created using the fixed value for the calibration parameters mentioned above.

This approach is not different from a standard Monte Carlo Markov chain. However, it is important to note that each evaluation in a point of the parameter space is extremely expensive in computational time since it implies generating five DES Y3-like mocks and measuring $w(\theta)$ on them. Moreover, we are not interested in the posterior of the calibration parameters $M_1$ and $\Delta_{\rm LM}$, but rather on their best-fit values, since this is the only quantity needed to generate mocks close to the real measurements. A straightforward approach would be to use a simple $\chi^2$ minimization algorithm to go directly to the minimum of the $\chi^2$ function, but the generation of the mocks contains an intrinsic random component when assigning the position and properties of galaxies. This introduces an important stochastic behavior in our problem and makes unusable the standard minimization algorithms.

In this work, following \citet{calibration}, we have decided to use the differential evolution stochastic minimization algorithm first proposed by \citet{Rainer1997}). The essential idea of the algorithm is to use a population of candidate solutions. We first initialize the population using a Latin Hypercube sampling; then, iteratively, these candidate solutions are combined to generate a new population and the $\chi^2$ is evaluated at each position. In more detail, the distance between two random candidate solutions is used to displace the best candidate solution so far (minimum $\chi^2$). If the new candidates are better than the previous ones they are accepted and belong to the new population; otherwise they are discarded and the new population is completed with candidates from the old population. Note that, because of this, the size of the population remains constant. The process ends when the standard deviation of the $\chi^2$ values of the population is smaller than a given tolerance times the mean of the $\chi^2$ values. The best candidate of the population at the end of the process is the best-fit used to generate the final mocks.

\begin{figure}
    \begin{center}
        \includegraphics[width=0.45 \textwidth]{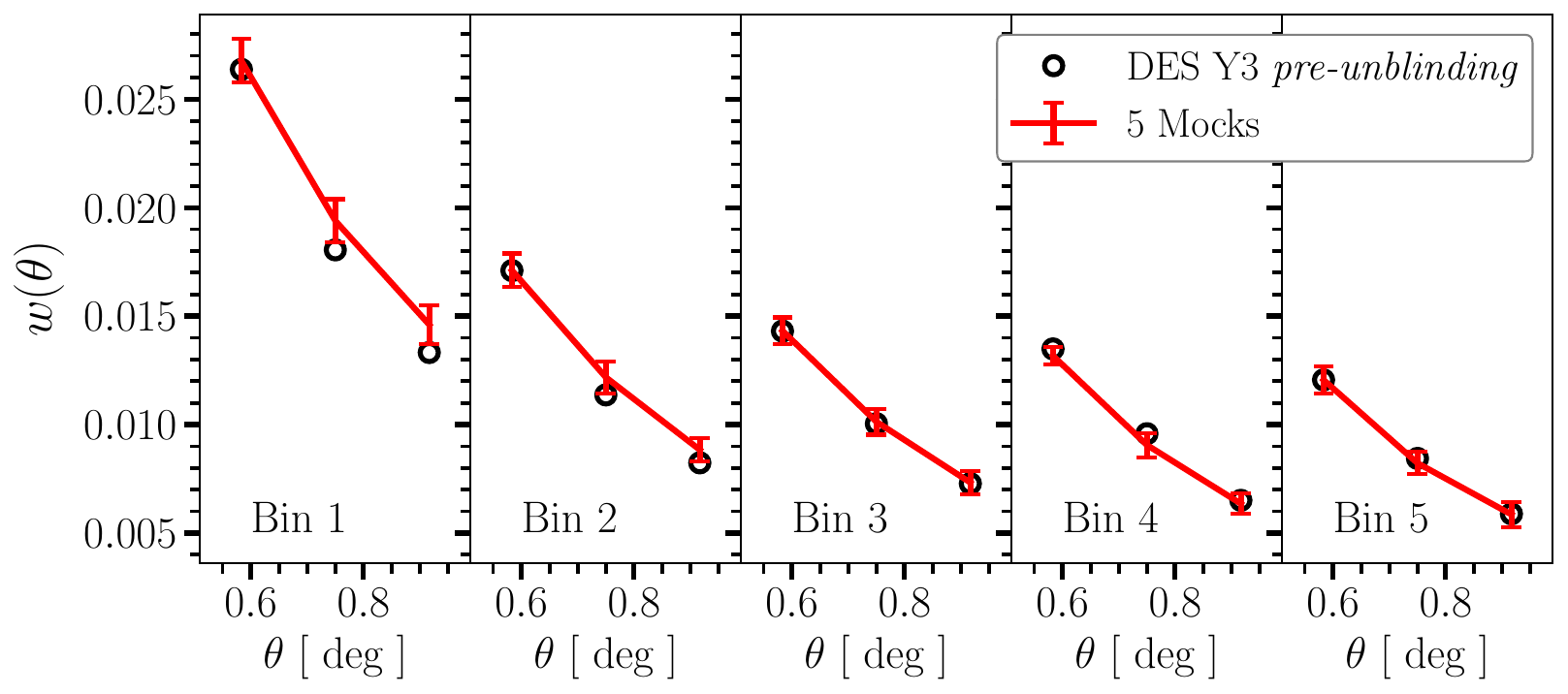}
    \end{center}
    \caption{Agreement between the \textit{pre-unblinding} $w(\theta)$ measurements of the data (open black circle) and the output of the calibrated mocks for the five tomographic redshift bins (red lines). The error bars from the mock measurements have been obtained as the square root of the diagonal of the covariance matrix using the 1952 mocks created with the fixed values.}
    \label{fig:calibration_agreement_data}
\end{figure}

Once the calibration parameters have been determined and the five mocks generated, we can check the agreement between the $w(\theta)$ from the mocks and the real measurements to verify how accurate the calibration is. The results are shown in Fig.~\ref{fig:calibration_agreement_data}. For each one of the tomographic bins, we represent the data with open black circles and the measurements from the five mocks with red lines. The errors have been obtained as the square root of the diagonal of the covariance matrix. The agreement is well within $1\sigma$ for all the bins, giving a goodness of the model $\chi_{\rm mod}^2/{\rm d.o.f} = 7.58/5$. Degrees of freedom (d.o.f.) equal five corresponds to the number of $\theta$-bins$=15$, three apertures times five tomographic bins, minus the ten free parameters, two (M$_1$ and $\Delta_{\rm LM}$) per bin.

%%%%%%%%%%%%%%%%%%%%%%%%%%%%%%%%%%%%%%%%%%%%%%%%%%%%%%%%%%%%%%%%%%%%%%%%%%%%%%%%%%%%%%%%%%%%%%%%%%%%%%%%%%%%%%%%%%%%%%%%%%%%%%%%%%%%
%----------------------------------------------------------------------------------------------------------------------------------
%%%%%%%%%%%%%%%%%%%%%%%%%%%%%%%%%%%%%%%%%%%%%%%%%%%%%%%%%%%%%%%%%%%%%%%%%%%%%%%%%%%%%%%%%%%%%%%%%%%%%%%%%%%%%%%%%%%%%%%%%%%%%%%%%%%%
\section{Analysis} \label{sec:resul} % used for referring to this section from elsewhere
In previous sections, we explained the methodology and the procedure used to create $1\,952$  mocks reproducing all the relevant properties of the DES Y3 \sample. In this section, we analyze the clustering of these mock catalogs in different spaces. We also include in the analysis a theoretical model prediction for those statistics.\\

\subsection{Theoretical model} \label{sec:tm} % used for referring to this section from elsewhere
The theoretical template is computed using the redshift-space power spectrum

\begin{align}
 P(k, \mu) &= ( 1 + \beta \mu^2 )^2 b^2 \big\{ [ P_{\rm lin}(k) - P_{\rm sm}(k) ] D_{\rm BAO} +  P_{\rm sm}(k) \big\} ,  
\end{align}
where $\mu$ is the dot product between $\hat{\boldsymbol{k}}$  and the line-of-sight direction, $b$ is the linear bias, and $\beta = f/ b $ with $f$ being the linear growth rate. The power spectrum is built using the linear power spectrum  $P_{\rm lin}(k)$ and the linear no-wiggle power spectrum $P_{\rm sm}$\footnote{Defined by following the 1D Gaussian smoothing in log-space described in Appendix A of \cite{2016JCAP...03..057V}.}. The nonlinear damping of the BAO feature is modeled by

\begin{equation}
D_{\rm BAO} (k,\mu)  = \exp\{ - k^2 [ \mu^2 \Sigma^2_{\shortparallel}  + (1-\mu^2) \Sigma^2_{\perp} + f \mu^2 (\mu^2 -1 ) \delta \Sigma^2 ]\} , 
\end{equation}    
with $\Sigma_{\shortparallel}  =  (1 + f) \Sigma_{\perp} $. The damping scales $\Sigma_\perp $  and  $\delta \Sigma $ are computed following \cite{Baldauf:2015xfa}. In MICE cosmology, $\Sigma_{\perp} = 5.80 \,$Mpc~$h^{-1}$ and $\delta \Sigma = 3.18  \,$Mpc~$h^{-1}$ at redshift 0 and they are scaled to higher redshift by the growth factor. See \cite{y3mainbao} for more details about the procedure to obtain these quantities.  Once provided with $P(k,\mu)$, we computed the anisotropic redshift-space correlation function $\xi(s,\mu)$ through a Fourier transform \citep[see][]{Chan2021}. The angular correlation function is obtained after projecting $\xi$ weighted by the redshift distribution $n(z)$  (normalized to 1),
.
\begin{equation}
\label{eq:w_template_schematic}
w(\theta) = \int dz_1\int dz_2 n(z_1) n(z_2)\xi \big( s(z_1,z_2,\theta),\mu(z_1,z_2,\theta) \big).
\end{equation} 

%The angular correlation function template $w$ is computed similar to \cite{CrocceCabreGazta_2011,Chan:2018gtc} but including the anisotropic damping of the BAO.  
The harmonic power spectrum template $C_\ell $ is derived from $w$ by a Legendre transform 

\begin{equation}
C_{\ell}=2 \pi \int_{-1}^{1} d \mu\  w(  \arccos (\mu ) ) \mathcal{L}_{\ell}(\mu), 
\end{equation}
where $ \mathcal{L}_\ell $ is the Legendre polynomial.  For more details on the modeling, see the main DES Y3 BAO paper \citep{y3mainbao}.

\subsection{Angular correlation function: $w(\theta)$} \label{sec:crs}

\begin{figure*}
    \begin{center}
        \includegraphics[width=\textwidth]{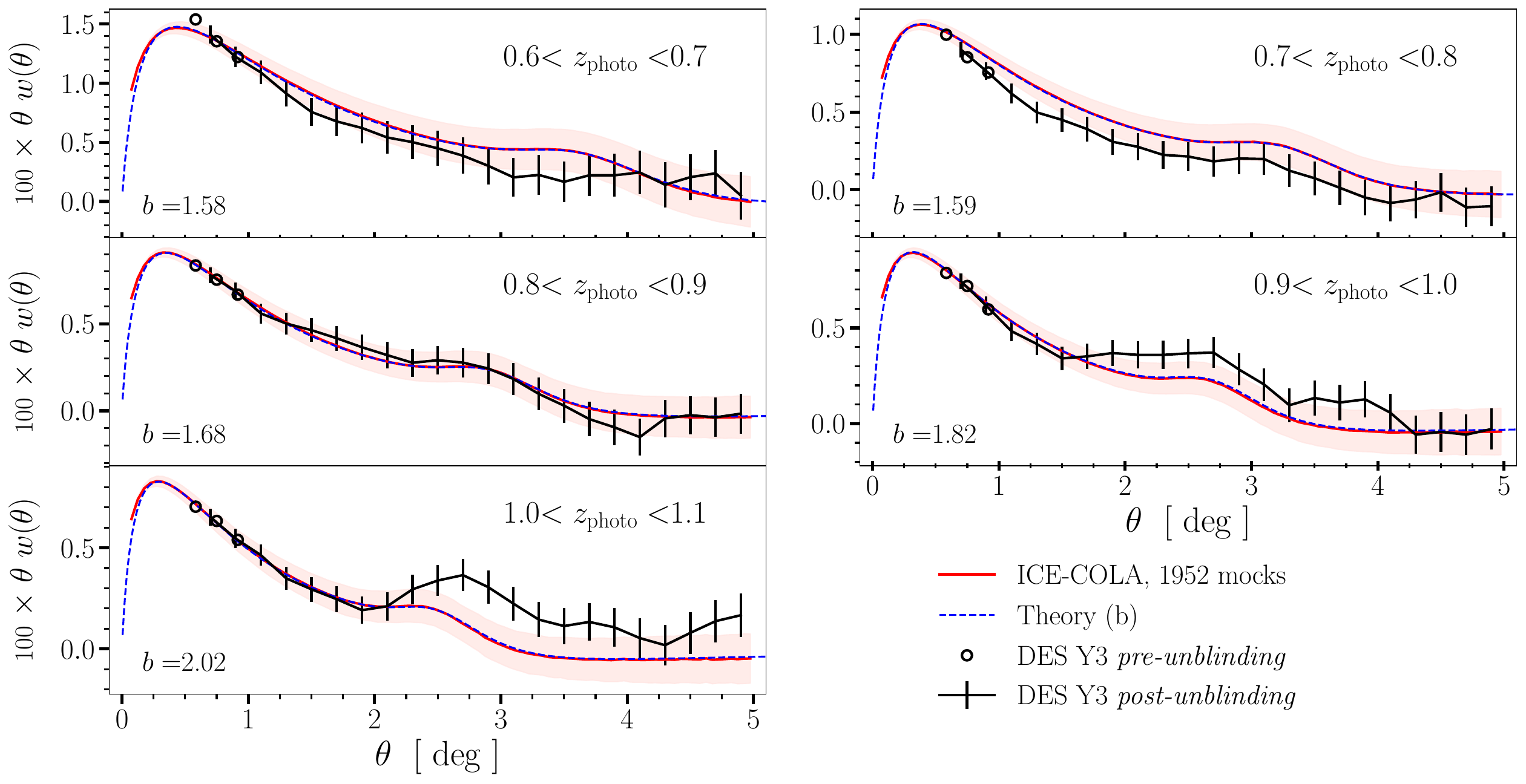}
    \end{center}
    \caption{Angular two-point correlation function for each tomographic bin. Red lines correspond to the average over all the mocks and shaded light-red bands correspond to the standard deviation. Dashed blue lines indicate the theoretical prediction described on Sec.~\ref{sec:tm} and black points correspond to the \textit{pre-unblinding} data values used in the calibration showed on Sec.~\ref{sec:Calib}. Finally, solid black lines with error bars represent the final \textit{post-unblinding} measurement of the data.}
    \label{fig:clustering}
\end{figure*}

Around ten thousand ACF must be calculated for the mocks ($1\,952$ mocks times five bins). For this reason, we resort to a code that allows the calculation using pixels, reducing the computational time. It is important to point out that since we only use angular apertures greater than one degree for fitting, any effect from the pixelization should be negligible. We use the public code CUTE \citep{Alonso2012}. CUTE supports the Landy \& Szalay estimator \citep{ls1993}:

\begin{equation}
    w(\theta) = \frac{DD(\theta) - 2DR(\theta)+RR(\theta)}{RR(\theta)},
\label{eq:lan-sza}
\end{equation}
where $DD$, $DR$ and $RR$ represent the total number of Data-Data, Data-Random and Random-Random pairs separated by an angular $\theta$ projected distance, respectively. In this case, Data correspond to galaxies in the mocks while Randoms are created by sampling the same volume with random points. The total number of randoms is $20$ times the average number of galaxies in the mocks. The same random catalog is used to calculate the clustering for all $1\,952$  mocks. The chosen pixel resolution is $n_{pix-shp}=4\,096$ which yields pixels with an angular $\theta$ resolution of $2.1$ arcmin.\\
Fig.~\ref{fig:clustering} shows the result of the angular two-point correlation function of $1\,952$  mocks for the 5 tomographic bins. It can be noticed that the difference between the \textit{pre-unblinding} values used for calibration (open black circles) and the averaged ACF for the mocks (solid red lines) differs from what was obtained on Sect.~\ref{sec:Calib} during the calibration procedure (see Fig.~\ref{fig:calibration_agreement_data}). However, these small differences are expected given the degree of representativeness when using only five mocks for the calibration ($0.25\%$ of the total number of mocks). It is important to keep in mind that the number of five mocks has been determined with a fixed calibration. Therefore, combining the error introduced by these fixed parameters and the allowed $2\%$ accuracy, we can expect the final accuracy with the best-fit calibration and all mocks to be slightly above $2\%$. To be more precise, this number has increased from $2$ to $3.1\%$. However, within the corresponding uncertainties, the agreement is still within $1\sigma$ for the five tomographic bins, giving a global $\chi_{\rm mod}^2/{\rm d.o.f} = 16.72/5$. The increase in the $\chi_{\rm mod}^2$ may be due either to the low representativeness of the five mocks used for the calibration or the presence of a strong anti-correlation among the data points. Nevertheless, when we quantify the goodness of the fit between the mocks to the data, using all $\theta$ bins in Fig.~\ref{fig:clustering}, we find remarkable good values $\chi_{m-d}^2/{\rm d.o.f} = [51.2,22.5,24.1,31.6,24.62]/22$. Only the first tomographic bin is away from having a $\chi_{\rm m-d}^2/{\rm d.o.f} \simeq 1$ and a bit less is the fourth bin. In conclusion, such accuracy is enough for our purposes and we do not consider rerunning the pipeline with more mocks in each point of the calibration. In more detail, only the first bin shows one \textit{pre-unblinding} data point out of $1\sigma$ from the mocks. In this case, the mean $w(\theta)$ of the mock has changed $\sim4\%$ from the value found during the calibration. Although this value is double of $2\%$ what was foreseen in the calibration (see Fig.~\ref{fig:area_number_mocks}), the global change is within the expectation. Blue curves in Fig.~\ref{fig:clustering} denotes the theoretical prediction described in Sect.~\ref{sec:tm}, and it is clear from the figure that modeled $w(\theta)$ agree almost perfectly with the measurements on the mocks. Finally, solid black lines with error bars correspond to the final \textit{post-unblinding} measurements of the data (using a brute force configuration of CUTE).\\

\subsection{Projected Clustering: $\xi_{w}(s_\perp)$} \label{sec:c3d} % used for referring to this section from elsewhere

In photometric surveys, most of the radial BAO information is lost due to redshift uncertainties. Additionally, the \photoz uncertainty causes the BAO scale in the 3D correlation function, $\xi(s)$, to deviate from its true position. However, \citet{Ross2017} demonstrated that when the correlation function is plotted against the transverse scale $ s_\perp = s \sqrt{ 1-\mu^2} $, the BAO peak appears where $ s_\perp $ equals to the true sound horizon scale. Thus, (angular) BAO information can still be retrieved via the 3D correlation analysis.    

\begin{figure}
\includegraphics[width=0.95\linewidth]{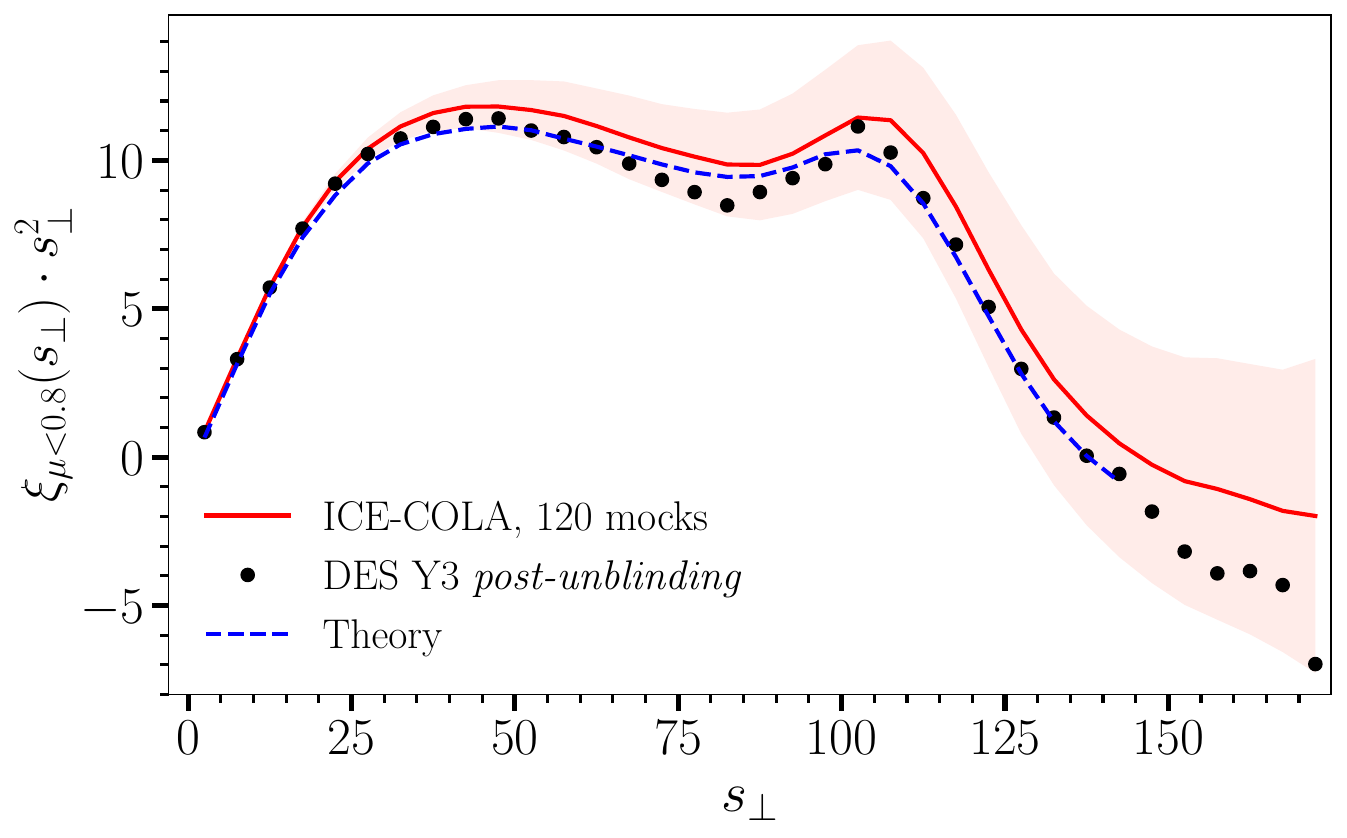}
\caption{3D wedge correlation function for $\mu$ range $[0,0.8]$.  With the same color code as previous figures, solid red line corresponds to ICE-COLA mocks and shaded light-red band to its standard deviation. \textit{Post-unblinding} measurement of data is shown with black points and the theory with a dashed blue line. }  
\label{fig:xiw_sperp_COLAmock}
\end{figure}

Following the methodology from \citet{Ross2017} and also described in \cite{y3mainbao}, we show in Fig.~\ref{fig:xiw_sperp_COLAmock} the 3D wedge correlation function $\xi_w(s_\perp) $ measured from the mocks (due to computational expenses, only $120$ mocks are used) and data.  
The results are also obtained from CUTE using Eq.~\eqref{eq:lan-sza}, by replacing $w(\theta)$ with $\xi(s_\perp,s_\parallel)$, and then integrating over $s_\parallel$ for the scales with $\mu<0.8$. %The setup of the measurements is similar to the $w$ case. 
As a comparison, we have also plotted the corresponding theory prediction. While \citet{Ross2017} assumes Gaussian \photoz distribution, the prediction makes use of the \photoz distribution from the mocks, resulting in better agreement with the numerical measurements. Further details of the comparison between the mock results and the theory will be presented in \citet{Chan2021}.
We find a good agreement for the 3D clustering of data, mocks and theory. 

\subsection{Angular power spectrum: $C_\ell$} \label{sec:chs} % used for referring to this section from elsewhere

We also measured the clustering signal in the Fourier conjugate space of angular distances on the sphere, the so-called harmonic space, by estimating the angular power spectra of galaxy number counts, $C_\ell$.
Although constructed from the same underlying field, the angular power spectrum and the correlation function present different advantages and downsides. Most notably, the correlation function is relatively straightforward to estimate in the presence of an angular survey mask. Still, its estimates are largely correlated. On the other hand, the power spectrum requires a deconvolution of the angular mask, but the correlation between scales reduces. Taking these pros and cons into account, it is clearly desirable to have complementary information from both statistics. See \cite{Giannantonio2016} for the first implementation of these complementary estimators in the context of DES Y1 data analyses.

We begin by estimating pixelized galaxy overdensity maps,
$\delta_g(\nv) = N_g(\nv) / \bar{N}_g - 1$,
for the ICE-COLA mocks, where $\nv$ is the pixel position in the sphere, $N_g(\nv)$ the pixelized galaxy number counts and $\bar{N}_g$ the mean number of galaxies per pixel.
We then used the ``Pseudo-$C_\ell$'' method~\citep{2002ApJ...567....2H} to estimate the angular power spectra.
The Pseudo-$C_\ell$ method deconvolves the incomplete sky coverage mode mixing effect on a set of band power bins using analytical methods and has the advantage of being less computationally expensive, reaching equivalent error estimates than optimal quadratic estimators. Also, Pseudo-$C_\ell$ estimators are effectively ``unbiased'' concerning maximum likelihood estimators.
In particular, we use the implementation of the {\tt NaMASTER} code\footnote{\url{https://github.com/LSSTDESC/NaMaster}}~\citep{2019MNRAS.484.4127A}.

The discrete nature of galaxy number counts introduces a shot-noise contribution to the estimated galaxy overdensity maps and, consequently, a bias to the estimated $C_\ell$.
We account for this ``noise bias'' analytically, following \citet{2019MNRAS.484.4127A} and \citet{2020JCAP...03..044N} by subtracting this Poissonian noise from our power spectrum.
For each ICE-COLA mock, we consider the partial sky coverage introduced by its associated mask, as shown in Fig.~\ref{fig:Masking}.
To optimize the BAO feature detection, we bin the power spectra in bands of $\Delta\ell=20$ from a minimum multipole $\ell_{\rm min}=10$ up to a maximum multipole of $1\,000$.

\begin{figure*}
\begin{center}
\includegraphics[width=\textwidth]{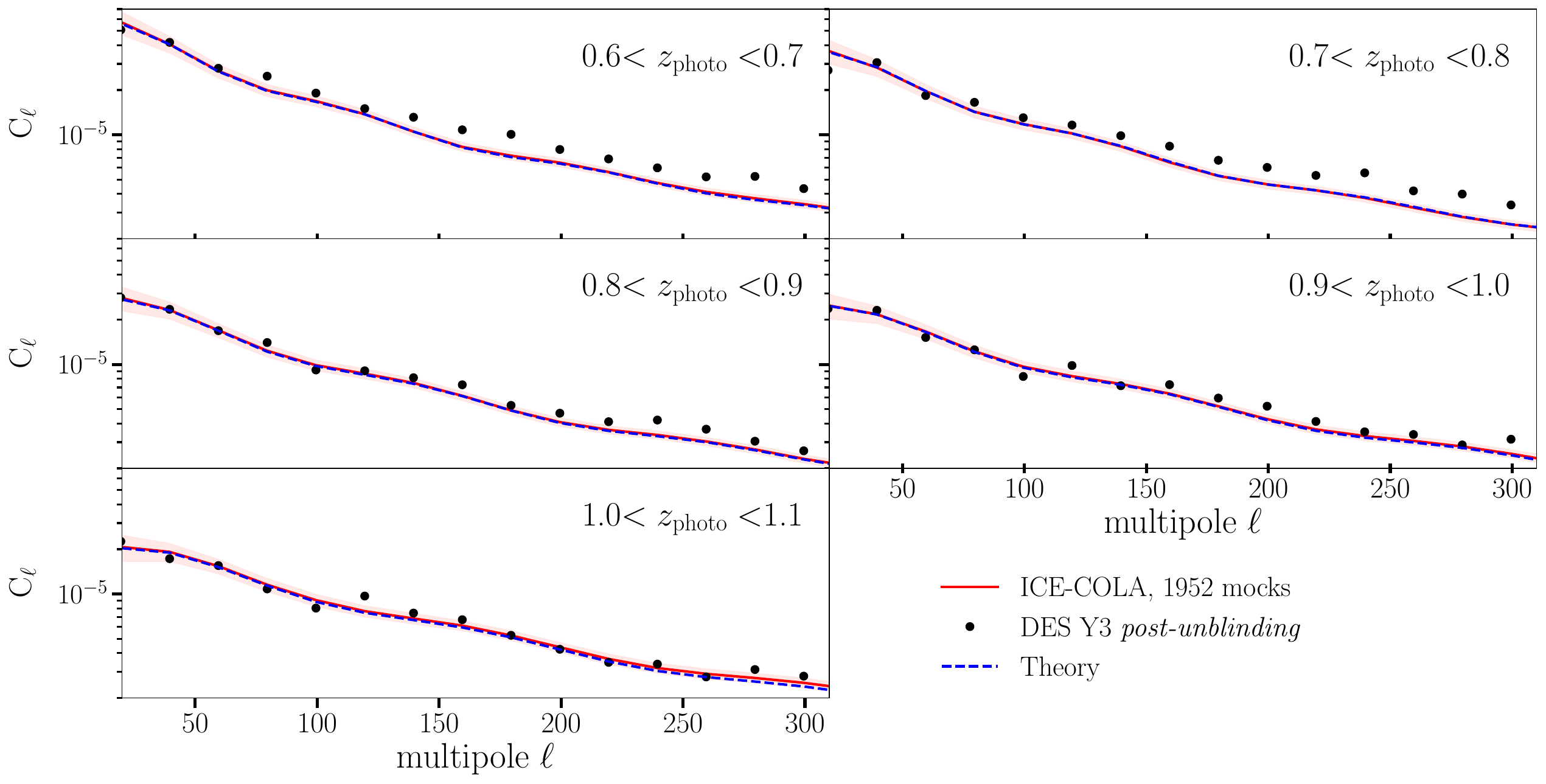}
\end{center}
\caption{Measured angular power spectra for the tomographic redshift bins considered.
Red lines correspond to the average over all the mocks and black points represent the \textit{post-unblinding} measurements of the data. 
The theoretical prediction is shown with dashed blue lines.
}
\label{fig:clustering_hs}
\end{figure*}

Figure~\ref{fig:clustering_hs} shows the mean and standard deviation of the estimated angular power spectra of the $1\,952$  ICE-COLA mocks
for the five tomographic bins (solid red lines) together with the theory prediction (dashed blue lines).
Black points correspond to the \textit{post-unblinding} measurements of the data.

\section{Effect of replications} \label{sec:rep} % used for referring to this section from elsewhere
Without taking into account the tails on redshift distribution for the tomographic bins on the edges ($[0.6-0.7]$ and $[1.0-1.1]$), the maximum comoving line-of-sight distance that can be found among galaxies on the mocks is $\sim1\,000$~Mpc~$h^{-1}$. Even if we consider the tails, the distance is lower than the box of the simulation, $1\,536$~Mpc~$h^{-1}$. In other words, no halo of ICE-COLA simulations is used more than one time along any give line-of-sight. But this problem does occur for different lines-of-sight. The area of the survey is very wide implying that for higher redshift several numbers of simulated boxes are used to equal the volume of the BAO Y3 \sample. Inevitably this implies the use of the same halo structures on each mock.  These repeated halos leave at different times but originate from the same initial structure they will obviously correlate, to leading order overdensities just grow linearly.

Table \ref{tab:repl} shows the upper limit for the percentage of replicated ``halos'' that can be found among bins. To obtain these values, we randomly sampled a box with particles and then created the light cone by replicating this box. These numbers are representative of all the repeated ``halos'' in a light cone and not only those used to make up the DES Y3 \sample, which are a few percent of the total. This fact, together with the selection efficiency, makes that the values are shown on table \ref{tab:repl} stand for a very conservative bound. However, these numbers refer to the replications present in a single mock but each light cone is used to create four mocks (see Fig.~\ref{fig:Masking}). This inevitably will introduce replications among different mocks created with the same light cone in addition to those among the bins of a single mock. Numbers in parentheses in table \ref{tab:repl} correspond to the percentage of repeated random particles among the four mocks made from one light cone. For example, the pair of bins (1,3) has on average (mean among the four mocks) $11.9\%$ repeated random particles. This number increase to $30.2\%$ when considering the four mocks. The difference between the $11.9$ and the $30.2\%$ come from the particles which are not repeated in one mock but do in others.

\begin{table}
\label{tab:repl}
\begin{center}
\caption{Percentage of replications among tomographic bins}
\label{tab:repl}
\scriptsize
\resizebox{0.49\textwidth}{!}{%
\begin{tabular}{|c|c|c|c|c|c|}
\hline
\textbf{BIN} & \textbf{1} & \textbf{2} & \textbf{3} & \textbf{4} & \textbf{5} \\ \hline
\textbf{1}   & 3.9 (22.6)    & 9.7 (30.6)     & 11.9 (30.2)      & 15.9 (30.3)    & 17.8 (30.8)     \\ \hline
\textbf{2}   & -          & 5.9 (27.9)      & 13.9 (28.7)      & 15.7 (29.1)     & 17.7 (28.9)     \\ \hline
\textbf{3}   & -          & -          & 8.1 (28.7)      & 16.5 (29.2)      & 18.5 (28.1)    \\ \hline
\textbf{4}   & -          & -          & -          & 8.9 (30.7)      & 18.2 (27.4)     \\ \hline
\textbf{5}   & -          & -          & -          & -          & 10.1 (29.9)      \\ \hline
\end{tabular}%
}
\end{center}
\smallskip
\footnotesize

Each value must be interpreted as the percentage of random points found duplicated in the corresponding pair of bins only in one mock. Numbers in parentheses represent the same but considering the four mocks done with one light cone

\end{table}

\begin{figure}
\begin{center}
\includegraphics[width=0.45\textwidth]{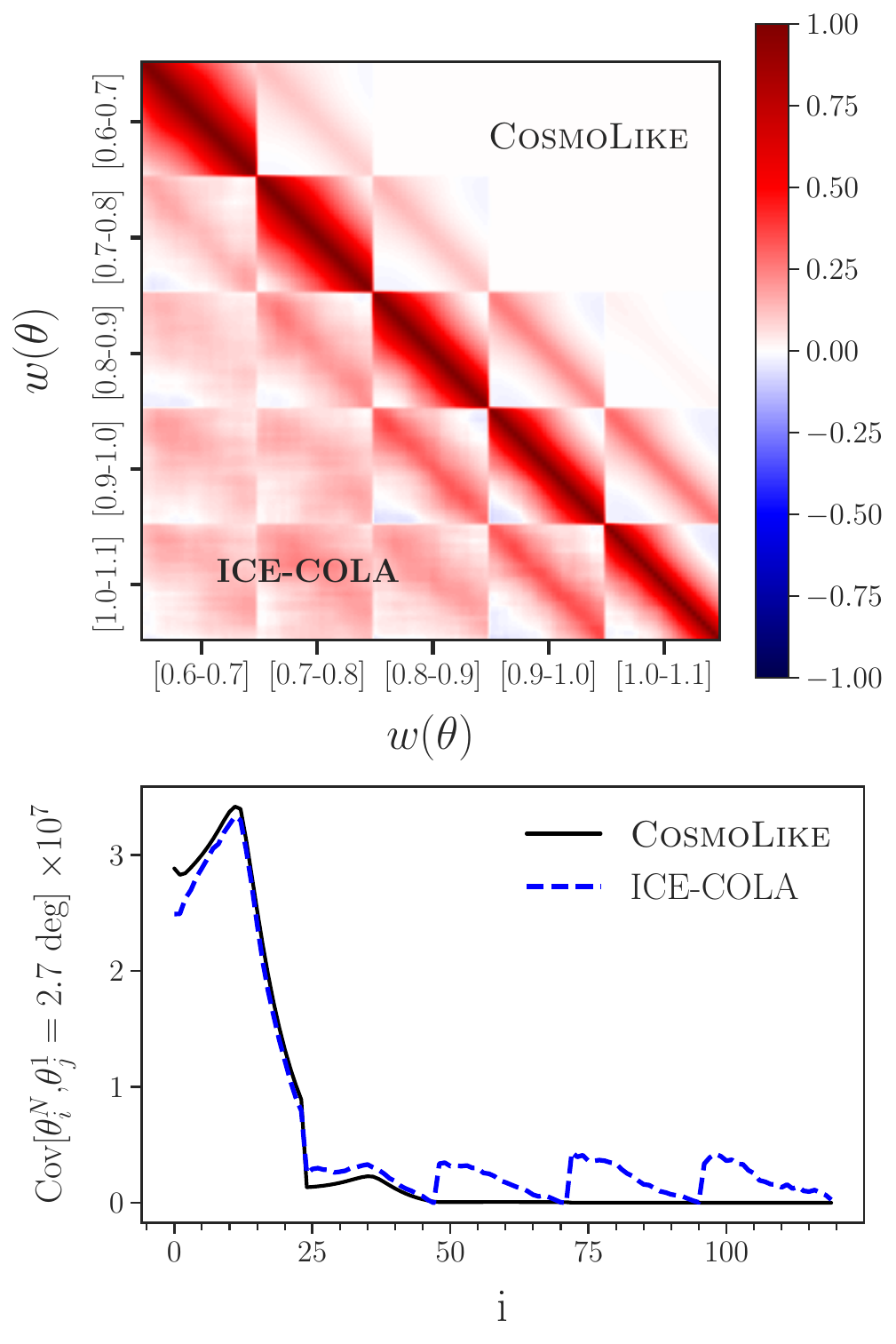}
\end{center}
\caption{Covariance matrix of the angular correlation function. Top: comparison of the correlation matrix obtained from \cosmolike halo model (upper diagonal) and from  ICE-COLA mocks (lower diagonal). Bottom: columm of the covariance matrix Cov[$\theta_{i}^{N},\theta_{j}^{M}$] with fixed $\theta_{j}^{1}=2.7$ deg. The $i$ and $j$ subscripts correspond to the index of tha data vector (from $0$ to $124$, with $\theta$ ranging in $[0,5]$deg in $\Delta\theta=0.2$deg steps and four redshift bins), and $N,M$ superscripts represent the index of the tomographic redshift bins.}
\label{fig:corrmatr}
\end{figure}

The main impact of this replication problem becomes noticeable when the cross-covariance matrix is analyzed. The repeated structures in different bins introduce a spurious correlation among measured $w(\theta)$ of tomographic bins. This effect is shown in Fig.~\ref{fig:corrmatr} wherein the top panel we compare the covariance matrix of the ICE-COLA mocks (lower diagonal) with the covariance matrix computed using \cosmolike halo model \citep[upper diagonal]{Cosmolike2017,cosmolike2}. From the former, the high degree of correlation between bins that are not adjacent is visible. This strong correlation can be seen clearly in the bottom panel of Fig.~\ref{fig:corrmatr} where we show one column of the covariance matrix corresponding to an aperture of $\theta=2.7$ deg for \cosmolike (solid black line) and ICE-COLA (dashed blue line). For simplicity, we are using in this plot a $\Delta \theta=0.2$.

%----------------------------------------------------------------------------------------------------
%%%%%%%%%%%%%%%%%%%%%%%%%%%%%%%%%%%%%%%%%%%%%%%%%%%%%%%%%%%%%%%%%%%%%%%%%%%%%%%%%%%%%%%%%%%%%%%%%%%%%%%%%%%%%%%%%%%%%%%%%%%%%%%%%%%%
\section{Conclusions} \label{sec:conclu} % used for referring to this section from elsewhere

The performance of well-validated mocks for DES Y3 BAO analysis is crucial for obtaining robust scientific results. The analysis of the data collected by the Dark Energy Survey during the first 3 years of the project poses a great scientific challenge in the development of the required mocks. We have created a significant number of mocks, $1\,952$, adequate for statistical analysis of galaxy clustering. The mocks have been built by populating the halo catalogs of $488$ ICE-COLA fast simulations. This task is made up of many steps that make up our pipeline, those are:

\begin{enumerate}
    \item HALO CATALOGS - Run $488$ ICE-COLA simulations with different initial conditions. Create light cones halo catalogs on-the-fly by replicating simulated boxes.\\
    \item GALAXY CATALOGS
    \begin{enumerate}
        \item HOD: Populate halos with one central galaxy and $N_{\rm sat}$ satellite galaxies following a Poisson distribution, setting the first free parameter per tomographic bin, $M_1$.
        \item HAM: Assign a pseudo-luminosity $l_{\rm p}$ to galaxies, setting the second free parameter per tomographic bin, $\Delta_{\rm LM}$.
        \item  Model photometric redshifts using a highly accurate method which follows a 2D probability distribution relating photometric and spectroscopic redshifts of VIPERS data.
        \item Apply four DES Y3 BAO mask over each full-sky light cone and quadrupling the total number of galaxy mocks compared to halo catalogs.
    \end{enumerate}
    \item CALIBRATION - Set an automatic calibration procedure using a differential evolution stochastic minimization algorithm to find the best values for the ten free parameters. This step is a loop of step (ii).\\
    \item FINAL GALAXY CATALOGS - Repeat step (ii) with the final set of parameters.
\end{enumerate}

The automatic calibration and the photometric redshift assignment procedure are key to obtaining final mocks that reproduce with high accuracy the principal properties of the data: (1) observed volume, (2) abundance of galaxies, redshift distribution, redshift uncertainty, and (3) clustering as a function of redshift. Firstly (1), the replications of the simulated box allowed us to achieve the observational volume of the DES Y3 \sample. However, the use of even bigger simulations would be beneficial, to avoid the repetition of structures and therefore the introduction of spurious cross-correlations and cross-covariance among not adjacent bins. This should be considered when designing mocks for the calculation of the covariance matrix in future surveys. Secondly (2), the use of overlapping galaxies between DES Y3 and VIPERS gave rise to a very realistic \photoz assignment. In these mocks, the abundance of galaxies, $n(z_{\rm photo})$, the redshift distribution $N(z_{\rm spec})$ and their uncertainties $\bar{z}$, $W68$ and $\sigma_{68}$ are in excellent agreement with VIPERS data, showing discrepancies lower than one per cent. Finally (3), the clustering measured in this set of mocks shows an excellent agreement with the \textit{pre-unblinding} data used for calibration. The uncertainty on the calibration procedure resulted in a goodness of the model $\chi_{\rm mod}^2/{\rm d.o.f} = 7.58/5$ while for the final mocks this value increased a bit up to $16.72/5$.\\
\cite{y3mainbao} presents the best fit and uncertainty for the angular scale of the BAO angular distance measurement $D_A$ using the Y3 Dark Energy Survey data release. The mocks presented in this work have been a crucial tool in the procedure of obtaining these results. ICE-COLA mocks have been used to run robustness tests and optimize the methodology. Finally, it is important to remark that the pipeline that created this set of mocks should work perfectly for any future surveys which need realistic mocks for galaxy clustering analysis.\\

%%%%%%%%%%%%%%%%%%%%%%%%%%%%%%%%%%%%%%%%%%%%%%%%%%%%%%%%%%%%%%%%%%%%%%%%%%%%%%%%%%%%%%%%%%%%%%%%%%%%%%%%%%%%%%%%%%%%%%%%%%%%%%%%%%%%
%----------------------------------------------------------------------------------------------------------------------------------
%%%%%%%%%%%%%%%%%%%%%%%%%%%%%%%%%%%%%%%%%%%%%%%%%%%%%%%%%%%%%%%%%%%%%%%%%%%%%%%%%%%%%%%%%%%%%%%%%%%%%%%%%%%%%%%%%%%%%%%%%%%%%%%%%%%%
\section*{Acknowledgements}
Funding for the DES Projects has been provided by the U.S. Department of Energy, the U.S. National Science Foundation, the Ministry of Science and Education of Spain, 
the Science and Technology Facilities Council of the United Kingdom, the Higher Education Funding Council for England, the National Center for Supercomputing 
Applications at the University of Illinois at Urbana-Champaign, the Kavli Institute of Cosmological Physics at the University of Chicago, 
the Center for Cosmology and Astro-Particle Physics at the Ohio State University,
the Mitchell Institute for Fundamental Physics and Astronomy at Texas A\&M University, Financiadora de Estudos e Projetos, 
Funda{\c c}{\~a}o Carlos Chagas Filho de Amparo {\`a} Pesquisa do Estado do Rio de Janeiro, Conselho Nacional de Desenvolvimento Cient{\'i}fico e Tecnol{\'o}gico and 
the Minist{\'e}rio da Ci{\^e}ncia, Tecnologia e Inova{\c c}{\~a}o, the Deutsche Forschungsgemeinschaft and the Collaborating Institutions in the Dark Energy Survey. 

The Collaborating Institutions are Argonne National Laboratory, the University of California at Santa Cruz, the University of Cambridge, Centro de Investigaciones Energ{\'e}ticas, 
Medioambientales y Tecnol{\'o}gicas-Madrid, the University of Chicago, University College London, the DES-Brazil Consortium, the University of Edinburgh, 
the Eidgen{\"o}ssische Technische Hochschule (ETH) Z{\"u}rich, 
Fermi National Accelerator Laboratory, the University of Illinois at Urbana-Champaign, the Institut de Ci{\`e}ncies de l'Espai (IEEC/CSIC), 
the Institut de F{\'i}sica d'Altes Energies, Lawrence Berkeley National Laboratory, the Ludwig-Maximilians Universit{\"a}t M{\"u}nchen and the associated Excellence Cluster Universe, 
the University of Michigan, NSF's NOIRLab, the University of Nottingham, The Ohio State University, the University of Pennsylvania, the University of Portsmouth, 
SLAC National Accelerator Laboratory, Stanford University, the University of Sussex, Texas A\&M University, and the OzDES Membership Consortium.

Based in part on observations at Cerro Tololo Inter-American Observatory at NSF's NOIRLab (NOIRLab Prop. ID 2012B-0001; PI: J. Frieman), which is managed by the Association of Universities for Research in Astronomy (AURA) under a cooperative agreement with the National Science Foundation.

The DES data management system is supported by the National Science Foundation under Grant Numbers AST-1138766 and AST-1\,536171.
The DES participants from Spanish institutions are partially supported by MICINN under grants ESP2017-89838, PGC2018-094773, PGC2018-102021, SEV-2016-0588, SEV-2016-0597, and MDM-2015-0509, some of which include ERDF funds from the European Union. IFAE is partially funded by the CERCA program of the Generalitat de Catalunya.
Research leading to these results has received funding from the European Research
Council under the European Union's Seventh Framework Program (FP7/2007-2013) including ERC grant agreements 240672, 291329, and 306478.
We  acknowledge support from the Brazilian Instituto Nacional de Ci\^encia
e Tecnologia (INCT) do e-Universo (CNPq grant 465376/2014-2).

This manuscript has been authored by Fermi Research Alliance, LLC under Contract No. DE-AC02-07CH11359 with the U.S. Department of Energy, Office of Science, Office of High Energy Physics.\\

This manuscript has been authored by Fermi Research Alliance, LLC under Contract No. DE-AC02-07CH11359 with the U.S. Department of Energy, Office of Science, Office of High Energy Physics. The simulation production and storage, as well as the processing and analysis tools have been developed, implemented and operated in collaboration with the Port d'Informaci\'{o} Cient\'{i}fica (PIC). SA was supported by the MICUES project, funded by the EU's H2020 MSCA grant agreement no. 713366 (InterTalentum UAM). ACR acknowledges financial support from the Spanish Ministry of Science, Innovation and Universities (MICIU) under grant AYA2017-84061-P, co-financed by FEDER (European Regional Development Funds) and by the Spanish Space Research Program ``Participation in the NISP instrument and preparation for the science of EUCLID" (ESP2017-84272-C2-1-R).

This project has received funding from the European Union's Horizon 2020 Research and Innovation Programme under the Marie Sk\l odowska-Curie grant agreement No 734374.\\
 
%%%%%%%%%%%%%%%%%%%%%%%%%%%%%%%%%%%%%%%%%%%%%%%%%%%%%%%%%%%%%%%%%%%%%%%%%%%%%%%%%%%%%%%%%%%%%%%%%%%%%%%%%%%%%%%%%%%%%%%%%%%%%%%%%%%%
%----------------------------------------------------------------------------------------------------------------------------------
%%%%%%%%%%%%%%%%%%%%%%%%%%%%%%%%%%%%%%%%%%%%%%%%%%%%%%%%%%%%%%%%%%%%%%%%%%%%%%%%%%%%%%%%%%%%%%%%%%%%%%%%%%%%%%%%%%%%%%%%%%%%%%%%%%%%
 
\bibliographystyle{aa}
\bibliography{main}

\label{lastpage}
\end{document}